\documentclass[	paper=a4, 
                fontsize=11pt, 
                footnotes=nomultiple,
                DIV=11
                ]{scrartcl}
\pdfoutput=1
\usepackage[]{amsmath}
\usepackage{amsthm}
\usepackage{pdfsync}
\usepackage{microtype}

\makeatletter
\@ifundefined{XeTeXversion}{
\usepackage[pdftex]{graphicx}
\usepackage[T1]{fontenc}
\usepackage{paratype}
\usepackage[charter,cal=cmcal]{mathdesign}
}{%
\usepackage[xetex]{graphicx}
\usepackage[no-math]{fontspec}%
 \setsansfont[Mapping=tex-text]{Calibri}
\setromanfont[Mapping=tex-text]{Cambria}
\usepackage{xunicode}%
\usepackage{xltxtra}%
\usepackage[xetex,
            colorlinks=true,
            linkcolor=spot,
            allcolors=spot,
            pdfauthor={Robin Ince},
            pdftitle={ }]{hyperref}
}
\makeatother
\graphicspath{{figs/}}
\usepackage[utf8]{inputenc}

\usepackage{setspace}
\recalctypearea

\usepackage[hyperref=true,
            backend=biber,
            bibencoding=utf8,
            style=authoryear,
            citestyle=authoryear,
            url=false,
            maxcitenames=2,
            maxbibnames=99,
            uniquename=false,
            isbn=false]{biblatex}

\DeclareFieldFormat{citehyperref}{%
  \DeclareFieldAlias{bibhyperref}{noformat}
  \bibhyperref{#1}}

\DeclareFieldFormat{textcitehyperref}{%
  \DeclareFieldAlias{bibhyperref}{noformat}
  \bibhyperref{%
    #1%
    \ifbool{cbx:parens}
      {\bibcloseparen\global\boolfalse{cbx:parens}}
      {}}}

\savebibmacro{cite}
\savebibmacro{textcite}

\renewbibmacro*{cite}{%
  \printtext[citehyperref]{%
    \restorebibmacro{cite}%
    \usebibmacro{cite}}}

\renewbibmacro*{textcite}{%
  \ifboolexpr{
    ( not test {\iffieldundef{prenote}} and
      test {\ifnumequal{\value{citecount}}{1}} )
    or
    ( not test {\iffieldundef{postnote}} and
      test {\ifnumequal{\value{citecount}}{\value{citetotal}}} )
  }
    {\DeclareFieldAlias{textcitehyperref}{noformat}}
    {}%
  \printtext[textcitehyperref]{%
    \restorebibmacro{textcite}%
    \usebibmacro{textcite}}}

\bibliography{fromzotero.bib}

\usepackage{chngpage}
\usepackage{url}
\usepackage[squaren]{SIunits}
\usepackage{color}
\definecolor{spot}{rgb}{0,0.2,0.6} 
\usepackage{marginnote}
 
\usepackage{bm}
\DeclareMathOperator{\sgn}{sgn}
\DeclareMathOperator*{\argmax}{arg\,max}
\usepackage{nicefrac}

\usepackage[format=plain,indention=.5cm,small,bf]{caption}

\usepackage[pdftex,
            colorlinks=true,
            allcolors=spot,
            pdfauthor={Robin Ince},
            pdftitle={ }]{hyperref}

\makeatletter
\DeclareOldFontCommand{\rm}{\normalfont\rmfamily}{\mathrm}
\DeclareOldFontCommand{\sf}{\normalfont\sffamily}{\mathsf}
\DeclareOldFontCommand{\tt}{\normalfont\ttfamily}{\mathtt}
\DeclareOldFontCommand{\bf}{\normalfont\bfseries}{\mathbf}
\DeclareOldFontCommand{\it}{\normalfont\itshape}{\mathit}
\DeclareOldFontCommand{\sl}{\normalfont\slshape}{\@nomath\sl}
\DeclareOldFontCommand{\sc}{\normalfont\scshape}{\@nomath\sc}
\makeatother

\theoremstyle{definition}
\newtheorem{defn}{Definition}[section]

\begin{document}

\author{Robin A. A. Ince\\
\textit{\small Institute of Neuroscience and Psychology}\\[-1.5mm]
\textit{\small University of Glasgow, UK}\\
{\normalsize\href{mailto:robin.ince@glasgow.ac.uk}{\texttt{robin.ince@glasgow.ac.uk}}}
}

\date{}

\title{Measuring multivariate redundant information with pointwise common change in surprisal}
\begin{singlespace}
    \maketitle
\end{singlespace}

\begin{abstract}
    The problem of how to properly quantify redundant information is an open question that has been the subject of much recent research.
    Redundant information refers to information about a target variable $S$ that is common to two or more predictor variables $X_i$.
    It can be thought of as quantifying overlapping information content or similarities in the representation of $S$ between the $X_i$.
    We present a new measure of redundancy which measures the common change in surprisal shared between variables at the local or pointwise level.
    We provide a game-theoretic operational definition of unique information, and use this to derive constraints which are used to obtain a maximum entropy distribution. 
    Redundancy is then calculated from this maximum entropy distribution by counting only those local co-information terms which admit an unambiguous interpretation as redundant information.
    We show how this redundancy measure can be used within the framework of the Partial Information Decomposition (PID) to give an intuitive decomposition of the multivariate mutual information into redundant, unique and synergistic contributions.
    We compare our new measure to existing approaches over a range of example systems, including continuous Gaussian variables.
    Matlab code for the measure is provided, including all considered examples.

\end{abstract}

\subsection*{Keywords}

Mutual information; Redundancy; Pointwise; Surprisal; Partial information decomposition

\section{Introduction}

Information theory was originally developed as a formal approach to the study of man-made communication systems \parencite{shannon_mathematical_1948,cover_elements_1991}.
However, it also provides a comprehensive statistical framework for practical data analysis \parencite{ince_statistical_2017}.
For example, mutual information is closely related to the log-likelihood ratio test of independence \parencite{sokal_biometry_1981}. 
Mutual information quantifies the statistical dependence between two (possibly multi-dimensional) variables.
When two variables ($X$ and $Y$) both convey mutual information about a third, $S$, this indicates that some prediction about the value of $S$ can be made after observing the values of $X$ and $Y$.
In other words, $S$ is represented in some way in $X$ and $Y$.
In many cases, it is interesting to ask how these two representations are related --- can the prediction of $S$ be improved by simultaneous observation of $X$ and $Y$ (synergistic representation), or is one alone sufficient to extract all the knowledge about $S$ which they convey together (redundant representation).
A principled method to quantify the detailed structure of such representational interactions between multiple variables would be a useful tool for addressing many scientific questions across a range of fields \parencite{timme_synergy_2013,williams_nonnegative_2010,wibral_partial_2017,lizier_framework_2014-1}. 
Within the experimental sciences, a practical implementation of such a method would allow analyses that are difficult or impossible with existing statistical methods, but that could provide important insights into the underlying system.

\textcite{williams_nonnegative_2010} present an elegant methodology to address this problem, with a non-negative decomposition of multivariate mutual information.
Their approach, called the Partial Information Decomposition (PID), considers the mutual information within a set of variables.
One variable is considered as a privileged \emph{target} variable, here denoted $S$, which can be thought of as the independent variable in classical statistics. 
The PID then considers the mutual information conveyed \emph{about} this target variable by the remaining \emph{predictor} variables, denoted $\mathcal{X} = \left\{X_1, X_2, \dots X_n\right\}$, which can be thought of as dependent variables.
In practice the target variable $S$ may be an experimental stimulus or parameter, while the predictor variables in $\mathcal{X}$ might be recorded neural responses or other experimental outcome measures. 
However, note that due to the symmetry of mutual information, the framework applies equally when considering a single (dependent) output in response to multiple inputs \parencite{wibral_partial_2017}.
\textcite{williams_nonnegative_2010} present a mathematical lattice structure to represent the set theoretic intersections of the mutual information of multiple variables \parencite{reza_introduction_1961}.
They use this to decompose the mutual information $I(\mathcal{X}; S)$ into terms quantifying the unique, redundant and synergistic information about the independent variable carried by each combination of dependent variables.
This gives a complete picture of the representational interactions in the system.

The foundation of the PID is a measure of redundancy between any collection of subsets of $\mathcal{X}$.
Intuitively, this should measure the information shared between all the considered variables, or alternatively their common representational overlap.
\textcite{williams_nonnegative_2010} use a redundancy measure they term $I_\text{min}$.
However as noted by several authors this measure quantifies the minimum \emph{amount} of information that all variables carry, but does not require that each variable is carrying the \emph{same} information.
It can therefore overstate the amount of redundancy in a particular set of variables.
Several studies have noted this point and suggested alternative approaches \parencite{griffith_quantifying_2014,harder_bivariate_2013,bertschinger_quantifying_2014,griffith_intersection_2014,bertschinger_shared_2013,olbrich_information_2015,griffith_quantifying_2015}.

In our view, the additivity of surprisal is the fundamental property of information theory that provides the possibility to meaningfully quantify redundancy, by allowing us to calculate overlapping information content.
In the context of the well-known set-theoretical interpretation of information theoretic quantities as measures which quantify the area of sets and which can be visualised with Venn diagrams \parencite{reza_introduction_1961}, co-information (often called interaction information) \parencite{mcgill_multivariate_1954,jakulin_quantifying_2003,bell_co-information_2003,matsuda_physical_2000} is a quantity which measures the intersection of multiple mutual information values (Figure \ref{fig:venn}).
However, as has been frequently noted, co-information conflates synergistic and redundant effects.

We first review co-information and the PID before presenting $I_\text{ccs}$, a new measure of redundancy based on quantifying the common change in surprisal between variables at the local or pointwise level \parencite{wibral_local_2014,lizier_local_2008,wibral_bits_2014,van_de_cruys_two_2011,church_word_1990}.
We provide a game-theoretic operational motivation for a set of constraints over which we calculate the maximum entropy distribution. 
This game-theoretic operational argument extends the decision theoretic operational argument of \textcite{bertschinger_quantifying_2014} but arrives at different conclusions about the fundamental nature of unique information.
We demonstrate the PID based on this new measure with several examples that have been previously considered in the literature.
Finally, we apply the new measure to continuous Gaussian variables \parencite{barrett_exploration_2015}.

\section{Interaction Information (co-Information)}

\subsection{Definitions}

The foundational quantity of information theory is \emph{entropy}, which is a measure of the variability or uncertainty of a probability distribution.
The entropy of a discrete random variable $X$, with probability mass function $P(X)$ is defined as:
\begin{align}
    H(X) = \sum_{x \in X} p(x) \log_2 \frac{1}{p(x)}
\end{align}
This is the expectation over $X$ of $h(x) = -\log_2 p(x)$, which is called the \emph{surprisal} of a particular value $x$.
If a value $x$ has a low probability, it has high surprisal and vice versa.
Many information theoretic quantities are similarly expressed as an expectation --- in such cases, the specific values of the function over which the expectation is taken are called \emph{pointwise} or \emph{local} values \parencite{wibral_local_2014,lizier_local_2008,wibral_bits_2014,van_de_cruys_two_2011,church_word_1990}.
We denote these local values with a lower case symbol\footnote{%
Note following \textcite{wibral_partial_2017} we denote probability denote probability distributions with a capital latter e.g. $P(X_1,X_2)$, but denote values of specific realisations, i.e. $P(X_1=x_1,X_2=x_2)$ with lower case shorthand $p(x_1,x_2)$.}.

\begin{figure}[htbp]
    \centering
    \includegraphics[width=0.8\textwidth]{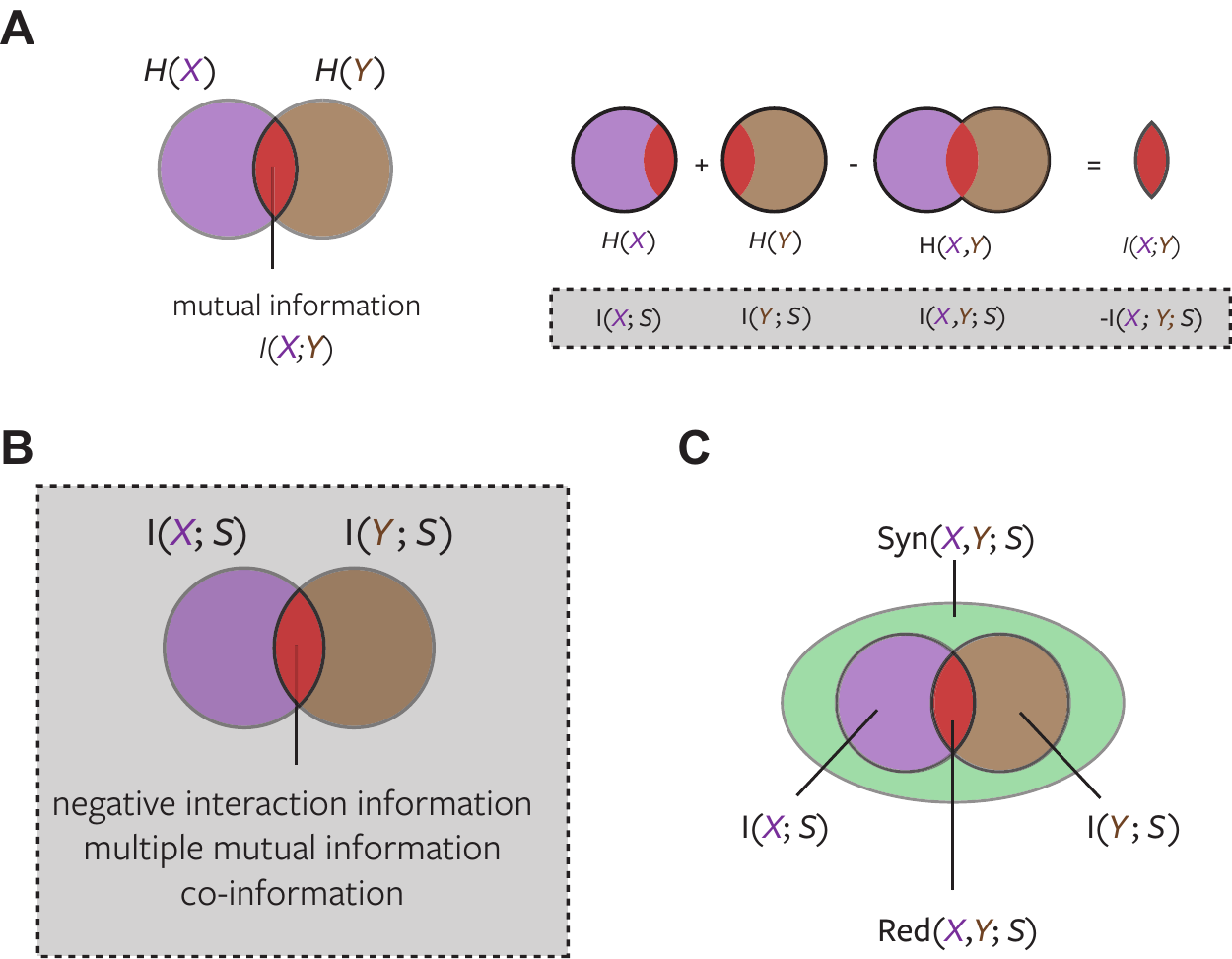}
    \caption{\emph{Venn diagrams of mutual information and interaction information.}
    \textbf{A.} Illustration of how mutual information is calculated as the overlap of two entropies. 
    \textbf{B.} The overlapping part of two mutual information values (negative interaction information) can be calculated in the same way (see dashed box in A).
    \textbf{C.} The full structure of mutual information conveyed by two variables about a third should separate redundant and synergistic regions.}
    \label{fig:venn}
\end{figure}

Figure \ref{fig:venn}A shows a Venn diagram representing the entropy of two variables $X$ and $Y$.
One way to derive mutual information $I(X; Y)$ is as the intersection of the two entropies.
This intersection can be calculated directly by summing the individual entropies (which counts the overlapping region twice) and subtracting the joint entropy (which counts the overlapping region once).
This matches one of the standard forms of the definition of mutual information: 
\begin{align}
    I(X; Y) &= H(X) + H(Y) - H(X,Y) \\
    &= \sum_{x,y} p(x,y) \left[ \log_2 \frac{1}{p(y)} - \log_2 \frac{1}{p(y|x)} \right]
\end{align}
Mutual information is the expectation of $i(x;y) = h(y) - h(y|x) = \log_2 \frac{p(y|x)}{p(y)}$, the difference in surprisal of value $y$ when value $x$ is observed\footnote{%
$p(y|x)$ denotes the conditional probability of observing $Y=y$, given that $X=x$ has been observed. $p(y|x) = \nicefrac{p(y,x)}{p(x)}$}.
To emphasise this point we use a notation which makes explicit the fact that local information measures a change in surprisal
\begin{align}
    i(x;y) &= \Delta_y h(x) = h(x) - h(x|y) \\
           &= \Delta_x h(y) = h(y) - h(y|x)
\end{align}
Mutual information is non-negative, symmetric and equals zero if and only if the two variables are statistically independent\footnote{$p(x,y) = p(x)p(y) \forall x \in X, y \in Y$} \parencite{cover_elements_1991}.

A similar approach can be taken when considering mutual information about a target variable $S$ that is carried by two predictor variables $X$ and $Y$ (Figure \ref{fig:venn}B).
Again the overlapping region can be calculated directly by summing the two separate mutual information values and subtracting the joint information.
However, in this case the resulting quantity can be negative.
Positive values of the intersection represent a net redundant representation: $X$ and $Y$ share the same information about $S$.
Negative values represent a net synergistic representation: $X$ and $Y$ provide more information about $S$ together than they do individually.

In fact, this quantity was first defined as the negative of the intersection described above, and termed \emph{interaction information} \parencite{mcgill_multivariate_1954}:
\begin{align} \label{eq:intinfo}
\begin{split}
    I( X; Y; S) &= I(X,Y; S) - I(X; S) - I(Y; S)\\
    &= I(S; X | Y) - I(S; X) \\
    &= I(S; Y | X) - I(S; Y) \\
    &= I(X; Y | S) - I(X; Y)
\end{split}
\end{align}
The alternative equivalent formulations illustrate how the interaction information is symmetric in the three variables, and also represents for example, the information between $S$ and $X$ which is gained (synergy) or lost (redundancy) when $Y$ is fixed (conditioned out). 

This quantity has also been termed \emph{multiple mutual information} \parencite{han_multiple_1980}, \emph{co-information} \parencite{bell_co-information_2003}, \emph{higher-order mutual information} \parencite{matsuda_physical_2000} and \emph{synergy} \parencite{gawne_how_1993,panzeri_correlations_1999,brenner_synergy_2000,schneidman_synergy_2003}. 
Multiple mutual information and co-information use a different sign convention from interaction information\footnote{%
For odd numbers of variables (e.g. three $X_1,X_2,S$) co-information has the opposite sign to interaction information; positive values indicate net redundant overlap.}.
 
As for mutual information and conditional mutual information, the interaction information as defined above is an expectation over the joint probability distribution.
Expanding the definitions of mutual information in Eq. \ref{eq:intinfo} gives:
\begin{align}
    I(X; Y; S) &= \sum_{x,y,s} p(x,y,s) \log_2 \frac{p(x,y,s)p(x)p(y)p(s)}{p(x,y)p(x,s),p(y,s)} \\
    I(X; Y; S) &= \sum_{x,y,s} p(x,y,s) \left[ \log_2 \frac{p(s|x,y)}{p(s)} - \log_2 \frac{p(s|x)}{p(s)} - \log_2 \frac{p(s|y)}{p(s)} \right]
\end{align}
As before we can consider the local or pointwise function 
\begin{align}
    i(x; y; s) = \Delta_s h(x,y) - \Delta_s h(x) - \Delta_s h(y)
\end{align}
The negation of this value measures the overlap in the change of surprisal about $s$ between values $x$ and $y$ (Figure \ref{fig:venn}A).

It can be seen directly from the definitions above that in the three variable case the interaction information is bounded:
\begin{align} \label{eq:intinfobounds}
\begin{split}
    I( X; Y; S) &\geq - \min \left[ I(S; X), I(S; Y), I(X; Y) \right]  \\
    I( X; Y; S) &\leq \hphantom{-} \min \left[ I(S; X|Y), I(S; Y|X), I(X; Y|S) \right] 
\end{split}
\end{align}

We have introduced interaction information for three variables, from a perspective where one variable is privileged (independent variable) and we study interactions in the representation of that variable by the other two. 
However, as noted interaction information is symmetric in the arguments, and so we get the same result whichever variable is chosen to provide the analysed information content.

Interaction information is defined similarly for larger numbers of variables.
For example, with four variables, maintaining the perspective of one variable being privileged, the 3-way Venn diagram intersection of the mutual information terms again motivates the definition of interaction information:
\begin{align}
\begin{split}
    I(W; X; Y; S) = &-I(W; S) - I(X; S) - I(Y; S) \\
    &+ I(W,X; S) + I(W,Y; S) + I(Y,X; S) \\
    &-I(W,X,Y; S)
\end{split}
\end{align}
In the n-dimensional case the general expression for interaction information on a variable set $\mathcal{V} = \left\{ \mathcal{X}, S \right\}$ where $\mathcal{X}=\left\{ X_1, X_2, \dots, X_n \right\}$ is:
\begin{align}
\begin{split}
    I\left(\mathcal{V}\right) = - \sum_{\mathcal{T}\subseteq \mathcal{X}} (-1)^{ \left| \mathcal{T} \right|}
    I\left( \mathcal{T}; S \right)
\end{split}
\label{eq:intgeneric}
\end{align}
which is an alternating sum over all subsets $\mathcal{T}\subseteq \mathcal{X}$, where each $\mathcal{T}$ contains $\left| \mathcal{T} \right|$ elements of $\mathcal{X}$. 
The same expression applies at the local level, replacing $I$ with the pointwise $i$.
Dropping the privileged target $S$ an equivalent formulation of interaction information on a set of n-variables $\mathcal{X}=\left\{ X_1, X_2, \dots, X_n \right\}$ in terms of entropy is given by \parencite{ting_amount_1962,jakulin_quantifying_2003}:
\begin{align}
\begin{split}
    I(\mathcal{X}) = - \sum_{\mathcal{T}\subseteq \mathcal{X}} (-1)^{ \left| \mathcal{X} \right| - \left| \mathcal{T} \right|}
    H\left(\mathcal{T}\right)
\end{split}
\end{align}

\subsection{Interpretation}

We consider as above a three variable system with a target variable $S$ and two predictor variables $X,Y$, with both $X$ and $Y$ conveying information about $S$.
The concept of redundancy is related to whether the information conveyed by $X$ and that conveyed by $Y$ is \emph{the same} or \emph{different}.
Within a decoding (supervised classification) approach, the relationship between the variables is determined from predictive performance within a cross-validation framework \parencite{quian_quiroga_extracting_2009,hastie_elements_2001}.
If the performance when decoding $X$ and $Y$ together is the same as the performance when considering e.g. $X$ alone, this indicates that the information in $Y$ is completely redundant with that in $X$; adding observation of $Y$ has no predictive benefit for an observer.
In practice redundancy may not be complete as in this example; some part of the information in $X$ and $Y$ might be shared, while both variables also convey unique information not available in the other.

The concept of synergy is related to whether $X$ and $Y$ convey more information when observed together than they do when observed independently.
Within the decoding framework this means higher performance is obtained by a decoder which predicts on a joint model of simultaneous $X$ and $Y$ observations, versus a decoder which combines independent predictions obtained from $X$ and $Y$ individually.
The predictive decoding framework provides a useful intuition for the concepts, but has problems quantifying redundancy and synergy in a meaningful way because of the difficulty of quantitatively relating performance metrics (percent correct, area under ROC, etc.) between different sets of variables --- i.e. $X$, $Y$ and the joint variable $(X,Y)$.

The first definition (Eq.~\ref{eq:intinfo}) shows that interaction information is the natural information theoretic approach to this problem: it contrasts the information available in the joint response to the information available in each individual response (and similarly obtains the intersection of the multivariate mutual information in higher order cases).
A negative value of interaction information quantifies the redundant overlap of Figure \ref{fig:venn}B, positive values indicate a net synergistic effect between the two variables.
However, there is a major issue which complicates this interpretation: interaction information conflates synergy and redundancy in a single quantity (Figure \ref{fig:venn}B) and so does not provide a mechanism for separating synergistic and redundant information (Figure \ref{fig:venn}C) \parencite{williams_nonnegative_2010}.
This problem arises for two reasons.
First, local terms $i(x;y;s)$ can be positive for some values of $x,y,s$  and negative for others.
These opposite effects can then cancel in the overall expectation.
Second, as we will see, the computation of interaction information can include terms which do not have a clear interpretation in terms of synergy or redundancy.

\section{The Partial Information Decomposition}
\label{sec:pid}

\begin{figure}[htbp]
    \centering
    \includegraphics[width=0.8\textwidth]{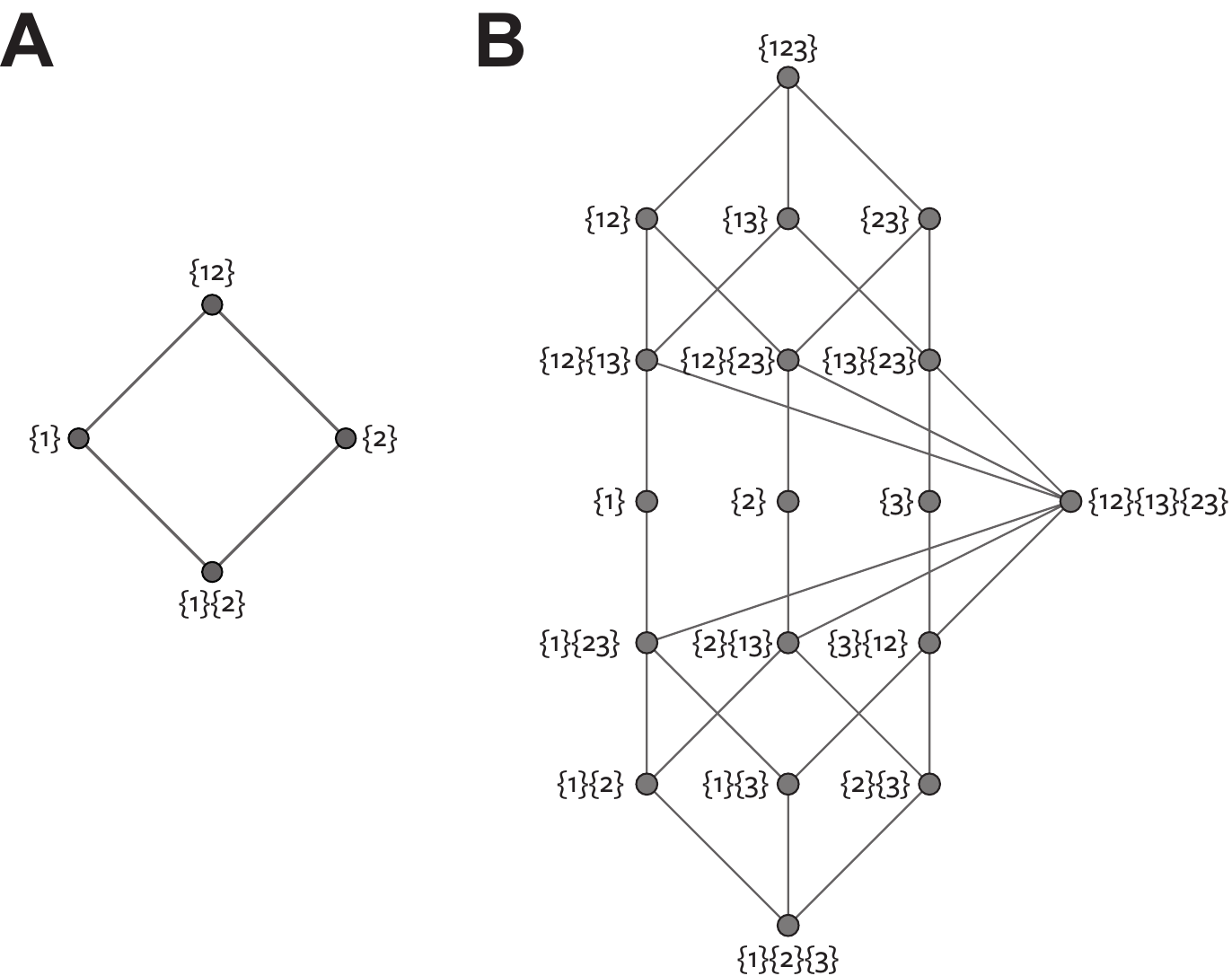}
    \caption{\emph{Redundancy lattice} for 
    \textbf{A.} two variables, 
    \textbf{B.} three variables. 
    Modified from \textcite{williams_nonnegative_2010}.}
    \label{fig:lattice}
\end{figure}

In order to address the problem of interaction information conflating synergistic and redundant effects, \textcite{williams_nonnegative_2010} proposed a decomposition of mutual information conveyed by a set of predictor variables $\mathcal{X}=\left\{ X_1, X_2, \dots, X_n \right\}$, about a target variable $S$. 
They reduce the total multivariate mutual information, $I(\mathcal{X}; S)$, into a number of non-negative atoms representing the unique, redundant and synergistic information between all subsets of $\mathcal{X}$: in the two-variable case this corresponds to the four regions of Figure \ref{fig:venn}C.
To do this they consider all subsets of $\mathcal{X}$, denoted $\mathbf{A_i}$, and termed \emph{sources}.
They show that the redundancy structure of the multi-variate information is determined by the ``collection of all sets of sources such that no source is a superset of any other'' --- formally the set of anti-chains on the lattice formed from the power set of $\mathcal{X}$ under set inclusion, denoted $\mathcal{A}(\mathcal{X})$.
Together with a natural ordering, this defines a redundancy lattice \parencite{crampton_completion_2001}.
Each node of the lattice represents a partial information atom, the value of which is given by a partial information (PI) function.
Note there is a direct correspondence between the lattice structure and a Venn diagram representing multiple mutual information values.
Each node on a lattice corresponds to a particular intersecting region in the Venn diagram. 
For two variables there are only four terms, but the advantage of the lattice representation becomes clearer for higher number of variables.
The lattice view is much easier to interpret when there are a large number of intersecting regions that are hard to visualise in a Venn diagram.
Figure \ref{fig:lattice} shows the structure of this lattice for $n=2,3$.
The PI value for each node, denoted $I_\partial$, can be determined via a recursive relationship (M\"obius inverse) over the redundancy values of the lattice:
\begin{align}
    I_\partial(S; \alpha) = I_\cap (S; \alpha) - \sum_{\beta \prec \alpha} I_\partial (S; \beta)
\end{align}
where $\alpha \in \mathcal{A}(\mathcal{X})$ is a set of sources (each a set of input variables $X_i$) defining the node in question.

The redundancy value of each node of the lattice, $I_\cap$, measures the total amount of redundant information shared between the sources included in that node.
For example, $I_\cap(S; \{X_1\},\{X_2\})$ quantifies the redundant information content about $S$ that is common to both $X_1$ and $X_2$.
The partial information function, $I_\partial$, measures the unique information contributed by only that node (redundant, synergistic or unique information within subsets of variables).

For the two variable case, if the redundancy function used for a set of sources is denoted $I_\cap\left(S; \mathbf{A_1},\dots,\mathbf{A_k}\right)$ and following the notation in \textcite{williams_nonnegative_2010}, the nodes of the lattice, their redundancy and their partial information values are given in Table \ref{tab:2varterms}.

\begin{table}[htbp]
\begin{center}
    \begin{tabular}{| c | p{2.7cm} | p{4cm} | p{4cm} |}
        \hline
        \textbf{Node label} & \textbf{Redundancy \newline function} & \textbf{Partial information} & \textbf{Represented atom} \\
        \hline \hline
        \{12\} & $I_\cap(S; \{ X_1, X_2 \} )$ & $I_\cap(S; \{ X_1, X_2 \} )$ \newline -$I_\cap(S; \{ X_1 \} )$ - $I_\cap(S;\{ X_2 \} )$ \newline +$I_\cap(S;\{ X_1 \}, \{ X_2 \} )$   & unique information in \newline $X_1$ and $X_2$ together (synergy) \\ \hline
        \{1\} & $I_\cap(S; \{ X_1 \} )$ & $I_\cap(S; \{ X_1 \} )$ \newline - $I_\cap(S; \{ X_1 \}, \{ X_2 \} )$ &  unique information in $X_1$ only \\ \hline
        \{2\} & $I_\cap(S; \{ X_2 \} )$ & $I_\cap(S; \{ X_2 \} )$ \newline - $I_\cap(S; \{ X_1 \}, \{ X_2 \} )$ & unique information in $X_2$ only \\ \hline
        \{1\}\{2\} & $I_\cap(S;\{ X_1 \}, \{ X_2 \} )$ & $I_\cap(S; \{ X_1 \}, \{ X_2 \} )$ & redundant information \newline between $X_1$ and $X_2$ \\ \hline
  \end{tabular}
  \caption{\emph{Full PID in the two-variable case.} The four terms here correspond to the four regions in Figure \ref{fig:venn}C.}
  \label{tab:2varterms}
\end{center}
\end{table}

Note that we have not yet specified a redundancy function.
A number of axioms have been proposed for any candidate redundancy measure \parencite{williams_nonnegative_2010,harder_bivariate_2013}:

\vspace{0.3cm}
\textbf{Symmetry:} 
\begin{equation}
    I_\cap\left(S; \mathbf{A_1},\dots,\mathbf{A_k} \right) \text{is symmetric with respect to the } \mathbf{A_i}\text{'s.}
\end{equation}

\textbf{Self Redundancy:} 
\begin{equation}
    I_\cap\left(S;\mathbf{A} \right) = I(S; \mathbf{A})
\end{equation}

\textbf{Subset Equality:} 
\begin{equation}
    I_\cap\left(S;\mathbf{A_1},\dots,\mathbf{A_{k-1}},\mathbf{A_k} \right) = I_\cap\left(S;\mathbf{A_1},\dots,\mathbf{A_{k-1}} \right)
    \mbox{ if } \mathbf{A_{k-1}} \subseteq \mathbf{A_k}
\end{equation}

\textbf{Monotonicity:} 
\begin{equation}
    I_\cap\left(S;\mathbf{A_1},\dots,\mathbf{A_{k-1}},\mathbf{A_k} \right) \leq I_\cap\left(S;\mathbf{A_1},\dots,\mathbf{A_{k-1}} \right)
\end{equation}

Note that previous presentations of these axioms have included subset equality as part of the monotonicity axiom; we separate them here for reasons that will become clear later.
Subset equality allows the full power set of all combinations of sources to be reduced to only the anti-chains under set inclusion (the redundancy lattice).
Self redundancy ensures that the top node of the redundancy lattice, which contains a single source $\mathbf{A}=\mathcal{X}$, is equal to the full multivariate mutual information and therefore the lattice structure can be used to decompose that quantity.
Monotonicity ensures redundant information is increasing with the height of the lattice, and has been considered an important requirement that redundant information should satisfy.

Other authors have also proposed further properties and axioms for measures of redundancy \parencite{griffith_intersection_2014,bertschinger_shared_2013}.
In particular, \textcite{harder_bivariate_2013} propose an additional axiom regarding the redundancy between two sources about a variable constructed as a copy of those sources:

\vspace{0.3cm}
\textbf{Identity Property (Harder et al.):}
\begin{equation}
    I_\cap\left(\left[\mathbf{A_1},\mathbf{A_2}\right]; \mathbf{A_1},\mathbf{A_2}\right) = I(\mathbf{A_1}; \mathbf{A_2})
    \label{eq:harderidentity}
\end{equation}

In this manuscript we focus on redundant and synergistic mutual information.
However, the concepts of redundancy and synergy can also be applied directly to entropy  \parencite{ince_partial_2017}.
Redundant entropy is variation that is shared between two (or more) variables, synergistic entropy is additional uncertainty that arises when the variables are considered together, over and above what would be obtained if they were statistically independent\footnote{%
Note that since the global joint entropy quantity is maximised when the two variables are independent, redundant entropy is always greater than synergistic entropy \parencite{ince_partial_2017}. However, local synergistic entropy can still occur: consider negative local information terms, which by definition quantify a synergistic local contribution to the joint entropy sum since  $h(x,y)>h(x)+h(y)$. }.
A crucial insight that results from this point of view is that mutual information itself quantifies both redundant and synergistic entropy effects --- it is the difference between redundant and synergistic entropy across the two inputs \parencite{ince_partial_2017}.
With $H_\partial$ denoting redundant or synergistic partial entropy analogous to partial information we have:
\begin{equation}
    I(\mathbf{A_1};\mathbf{A_2}) = H_\partial(\{ \mathbf{A_1} \}\{ \mathbf{A_2} \}) - H_\partial(\{ \mathbf{A_1}\mathbf{A_2} \})
    \label{eq:miped}
\end{equation}
This is particularly relevant for the definition of the identity axiom. 
We argue that the previously unrecognised contribution of synergistic entropy to mutual information (pointwise negative terms in the mutual information expectation sum) should not be included in an information redundancy measure.

Note that any information redundancy function can induce an entropy redundancy function by considering the information redundancy with the copy of the inputs.
For example, for the bivariate case we can define:
\begin{equation}
    H_\cap( \{ \mathbf{A_1}\} \{\mathbf{A_2}\} ) = I_\cap\left(\left[\mathbf{A_1},\mathbf{A_2}\right]; \mathbf{A_1},\mathbf{A_2}\right) 
\end{equation}
So any information redundancy measure that satisfies the identity property \parencite{bertschinger_quantifying_2014} cannot measure synergistic entropy \parencite{ince_partial_2017}, since for the induced entropy redundancy measure $H_\cap( \{ \mathbf{A_1}\} \{\mathbf{A_2}\} ) =  I(\mathbf{A_1}; \mathbf{A_2})$ so from Eq.~\ref{eq:miped} $H_\partial(\{ \mathbf{A_1}\mathbf{A_2} \})=0$.
To address this without requiring introducing in detail the partial entropy decomposition \parencite{ince_partial_2017}, we propose a modified version of the identity axiom, which still addresses the two-bit copy problem but avoids the problem of including synergistic mutual information contributions in the redundancy measure. 
When $I(\mathbf{A_1};\mathbf{A_2})=0$ there are no synergistic entropy effects because $i(a_1,a_2)=0 \; \forall a_1,a_2$ so there are no misinformation terms and no synergistic entropy between the two inputs. 

\vspace{0.3cm}
\textbf{Independent Identity Property:}
\begin{equation}
    I(\mathbf{A_1}; \mathbf{A_2})=0 \implies I_\cap\left(\left[\mathbf{A_1},\mathbf{A_2}\right]; \mathbf{A_1},\mathbf{A_2}\right) = 0
    \label{eq:indidentity}
\end{equation}

Please note that while this section primarily reviews existing work on the partial information decomposition, two novel contributions here are the explicit consideration of subset equality separate to monotonicity, and the definition of the independent identity property. 

\subsection{An example PID: RdnUnqXor}
\label{sec:examplerdnunqxor}

\begin{figure}[htbp]
    \centering
    \includegraphics[width=0.8\textwidth]{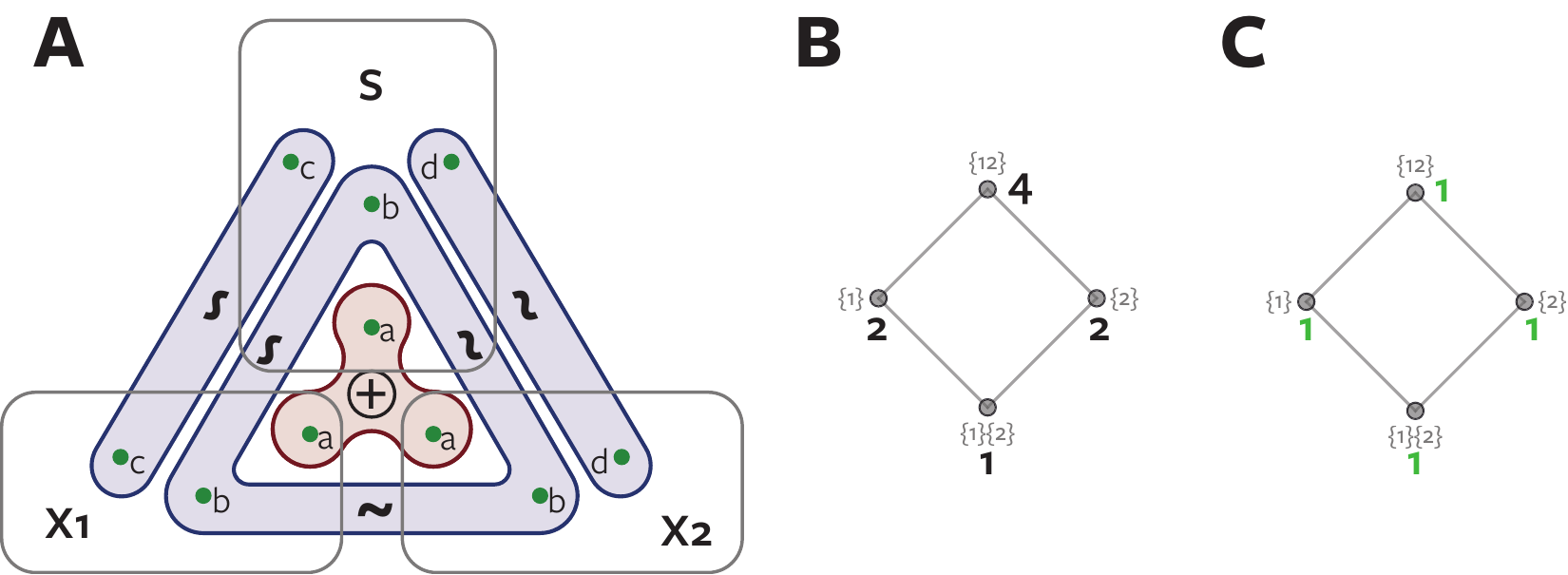}
    \caption{\emph{Partial Information Decomposition for \textsc{RdnUnqXor}}
    \textbf{A}: The structure of the \textsc{RdnUnqXor} system borrowing the graphical representation from \parencite{james_multivariate_2016}. $S$ is a variable containing 4 bits (labelled $a,b,c,d$). $X_1$ and $X_2$ each contain 3 bits. $\sim$ indicates bits which are coupled (distributed identically) and $\oplus$ indicates the enclosed variables form the \textsc{xor} relation.  
    \textbf{B}: Redundant information values on the lattice (black).
    \textbf{C}: Partial information values on the lattice (green).
    }
    \label{fig:rdnunqxor}
\end{figure}

Before considering specific measures of redundant information that have been proposed for use with the PID, we first illustrate the relationship between the redundancy and the partial information lattice values with an example.
We consider a system called \textsc{RdnUnqXor} \parencite{griffith_quantifying_2014}.
The structure of this system is shown in Figure~\ref{fig:rdnunqxor}A \parencite{james_multivariate_2016}. 
It consists of two three bit predictors, $X_1$ and $X_2$, and a four bit target $S$.
This example is noteworthy, because an intuitive PID is obvious from the definition of the system, and it includes by construction $1$ bit of each type of information decomposable with the PID.

All three variables share a bit (labelled \emph{b} in Fig.~\ref{fig:rdnunqxor}A).
This means there should be $1$ bit of redundant information.
Bit \emph{b} is shared between each predictor and the target so forms part of $I(X_i;S)$, and is also shared between the predictors, therefore it is shared or redundant information.
All variables have one bit that is distributed according to a \textsc{xor} configuration across the three variables (labelled \emph{a}). 
This provides $1$ bit of synergy within the system, because the value of bit \emph{a} of $S$ can only be predicted when $X_1$ and $X_2$ are observed together simultaneously \parencite{griffith_quantifying_2014}.
Bits \emph{c} and \emph{d} are shared between $S$ and each of $X_1$ and $X_2$ individually.
So each of these contributes to $I(X_i;S)$, but as unique information. 

We illustrate the calculation of the PID for this system (Figure~\ref{fig:rdnunqxor}B,C, Table~\ref{tab:rndunqxor}).
From the self-redundancy axiom, the three single-source terms can all be calculated directly from the classical mutual information values. 
The single predictors each have 2 bits of mutual information (the two bits shared with $S$).
Both predictors together have four bits of mutual information with $S$, since the values of all four bits of $S$ are all fully determined when both $X_1$ and $X_2$ are observed.
Since by construction there is $1$ bit shared redundantly between the predictors, we claim $I_\cap(S;\{1\}\{2\})=1$ bit and we have all the redundancy values on the lattice. 
Then from the summation procedure illustrated in Table~\ref{tab:2varterms} we can calculate the partial information values. 
For example, $I_\partial(S;\{1\}) = 2 - 1 = 1$, and $I_\partial(S;\{12\}) = 4 - 1 - 1 - 1 = 1$.

\begin{table}[htbp]
\begin{center}
\begin{tabular}{| c  | c | c |}
\hline
\textbf{Node} & $I_\cap$ & ${I_\partial}$     \\
\hline \hline
$\{1\}\{2\}$  & $1$      & $1$                \\
$\{1\}$       & $2$      & $1$                \\
$\{2\}$       & $2$      & $1$                \\
$\{12\}$      & $4$      & $1$        \\
\hline
\end{tabular}
\caption{\emph{PID for \textsc{RdnUnqXor} (Figure~\ref{fig:rdnunqxor}).}}
\label{tab:rndunqxor}
\end{center}
\end{table}

\subsection{Measuring redundancy with minimal specific information: \texorpdfstring{$\bm{I}_\text{min}$}{Imin}}
\label{sec:imin}

The redundancy measure proposed by \textcite{williams_nonnegative_2010} is denoted $I_\text{min}$ and derived as the average (over values $s$ of $S$) minimum specific information \parencite{deweese_how_1999,butts_how_2003} over the considered input sources.
The information provided by a source $\mathbf{A}$ (as above a subset of dependent variables $X_i$) can be written:
\begin{equation}
    I(S; \mathbf{A}) = \sum_s p(s)I(S=s; \mathbf{A})
\end{equation}
where $I(S=s; \mathbf{A})$ is the \emph{specific information}:
\begin{equation}
    I(S=s; \mathbf{A}) = \sum_{\mathbf{a}} p(\mathbf{a}|s) \left[ \log_2 \frac{1}{p(s)} - \log_2 \frac{1}{p(s|\mathbf{a})} \right]
\end{equation}
which quantifies the average reduction in surprisal of $s$ given knowledge of $\mathbf{A}$.
This splits the overall mutual information into the reduction in uncertainty about each individual target value.
$I_\text{min}$ is then defined as:
\begin{equation}
    I_\text{min}(S; \mathbf{A_1},\dots,\mathbf{A_k} ) = \sum_s p(s) \min_\mathbf{A_i} I(S=s; \mathbf{A_i})
    \label{eq:imin}
\end{equation}

This quantity is the expectation (over $S$) of the minimum amount of information about each specific target value $s$ conveyed by any considered source.
$I_\text{min}$ is non-negative and satisfies the axioms of symmetry, self redundancy and monotonicity, but not the identity property (neither Harder et al.~or independent forms).
The crucial conceptual problem with $I_\text{min}$ is that it indicates the variables share a common \emph{amount} of information, but not that they actually share the \emph{same} information content \parencite{harder_bivariate_2013,timme_synergy_2013,griffith_quantifying_2014}.

The most direct example of this is the ``two-bit copy problem'', which motivated the identity axiom \parencite{harder_bivariate_2013,timme_synergy_2013,griffith_quantifying_2014}.
We consider two independent uniform binary variables $X_1$ and $X_2$ and define $S$ as a direct copy of these two variables $S=(X_1,X_2)$.
In this case $I_\text{min}(S; \{1\}\{2\})=1$ bit; for every $s$ both $X_1$ and $X_2$ each provide $1$ bit of specific information.
However, both variables give different information about each value of $s$: $X_1$ specifies the first component, $X_2$ the second.
Since $X_1$ and $X_2$ are independent by construction there should be no overlap.
This illustrates that $I_\text{min}$ can overestimate redundancy with respect to an intuitive notion of overlapping information content.

\subsection{Measuring redundancy with maximised co-information: \texorpdfstring{$\bm{I}_\text{broja}$}{Ibroja}}
\label{sec:ibroja}

A number of alternative redundancy measures have been proposed for use with the PID in order to address the problems with $I_\text{min}$ \parencite[reviewed by][]{barrett_exploration_2015}.
Two groups have proposed an equivalent approach, based on the idea that redundancy should arise only from the marginal distributions $P(X_1,S)$ and $P(X_2,S)$ \parencite[Assumption *]{bertschinger_quantifying_2014}\footnote{Please note that we follow their terminology and refer to this concept as assumption (*) throughout.} and that synergy should arise from structure not present in those two marginals, but only in the full joint distribution $P(X_1,X_2,S)$.
\textcite{griffith_quantifying_2014} frame this view as a minimisation problem for the multivariate information $I(S;X_1,X_2)$ over the class of distributions which preserve the individual source-target marginal distributions.
\textcite{bertschinger_quantifying_2014} seek to minimise $I(S;X_1|X_2)$ over the same class of distributions, but as noted both approaches result in the same PID.
In both cases the redundancy, $I_\cap(S;\{X_1\}\{X_2\})$, is obtained as the maximum of the co-information (negative interaction information) over all distributions that preserve the source-target marginals:
\begin{align}
    I_\text{max-nii}(S;\{X_1\}\{X_2\}) &= \max_{Q \in \Delta_P} -I_Q(S; X_1; X_2) \\
    \Delta_P &= \left\{ Q \in \Delta : Q(X_1,S) = P(X_1,S), Q(X_2,S) = P(X_2,S) \right\}
    \label{eq:deltap}
\end{align}

We briefly highlight here a number of conceptual problems with this approach. 
First, this measure satisfies the Harder et al.~identity property (Eq.~\ref{eq:harderidentity}) \parencite{harder_bivariate_2013,bertschinger_quantifying_2014} and is therefore incompatible with the notion of synergistic entropy \parencite{ince_partial_2017}.
Second, this measure optimises co-information, a quantity which conflates synergy and redundancy \parencite{williams_nonnegative_2010}.
Given \parencite[Assumption *]{bertschinger_quantifying_2014} which states that unique and redundant information are constant on the optimisation space, this is equivalent to minimizing synergy \parencite{wibral_partial_2017}.
\begin{align}
    I_\text{max-nii}(S;\{X_1\}\{X_2\}) &= I_\text{red}(S;\{X_1\}\{X_2\}) - I_\text{syn-min}(S;\{X_1\}\{X_2\}) 
\end{align}
where $I_\text{syn-min}(S;\{X_1\}\{X_2\})$ is the smallest possible synergy given the target-predictor marginal constraints, but is not necessarily zero.
Therefore, the measure provides a bound on redundancy (under \parencite[Assumption *]{bertschinger_quantifying_2014}) but cannot measure the true value.
Third, \textcite{bertschinger_quantifying_2014} motivate the constraints for the optimisation from an operational definition of unique information based on decision theory. 
It is this argument which suggests that the unique information is constant on the optimisation space $\Delta_P$, and which motivates a foundational axiom for the measure that equal target-predictor marginal distributions imply zero unique information\footnote{%
``Consider the case $\mathcal{Y}=\mathcal{Z}$ and $\kappa=\mu \in K(\mathcal{X};\mathcal{Y})$, i.e. $Y$ and $Z$ use a similar channel. In this case $Y$ has no unique information with respect to $Z$, and $Z$ has no unique information with respect to $Y$.'' \parencite{bertschinger_quantifying_2014}}.
However, we do not agree that unique information is invariant to the predictor-predictor marginal distributions, or necessarily equals zero when target-predictor marginals are equal.
We revisit the operational definition in Section~\ref{sec:operational} by considering a game theoretic extension which provides a different perspective. 
We use this to provide a counter-example that proves the decision theoretic argument is not a necessary condition for the existence of unique information, and therefore the $I_\text{broja}$ procedure is invalid since redundancy is not fixed on $\Delta_P$.
We also demonstrate with several examples (Section~\ref{sec:examples}) how the $I_\text{broja}$ optimisation results in coupled predictor variables, suggesting the co-information optimisation is indeed maximising the source redundancy between them. 

\subsection{Other redundancy measures}

\textcite{harder_bivariate_2013} define a redundancy measure based on a geometric projection argument, which involves an optimisation over a scalar parameter $\lambda$, and is defined only for two sources, so can be used only for systems with two predictor variables.
\textcite{griffith_intersection_2014} suggest an alternative measure motivated by zero-error information, which again formulates an optimisation problem (here maximisation of mutual information) over a family of distributions (here distributions $Q$ which are a function of each predictor so that $H(Q|X_i)=0$). 
\textcite{griffith_quantifying_2015} extend this approach by modifying the optimisation constraint to be $H(Q|X_i)=H(Q|X_i,Y)$.

\section{Measuring redundancy with pointwise common change in surprisal: \texorpdfstring{$\bm{I}_\text{ccs}$}{Iccs}}

We derive here from first principles a measure that we believe encapsulates the intuitive meaning of redundancy between sets of variables.
We argue that the crucial feature which allows us to directly relate information content between sources is the additivity of surprisal. 
Since mutual information measures the expected change in pointwise surprisal of $s$ when $x$ is known, we propose measuring redundancy as the expected pointwise change in surprisal of $s$ which is common to $x$ and $y$.
We term this \emph{common change in surprisal} and denote the resulting measure $I_{\text{ccs}}(S; \alpha)$.

\subsection{Derivation}

As for entropy and mutual information we can consider a Venn diagram (Figure \ref{fig:venn}) for the change in surprisal of a specific value $s$ for specific values $x$ and $y$ and calculate the overlap directly using local co-information (negative local interaction information).
However, as noted before the interaction information can confuse synergistic and redundant effects, even at the pointwise level.
Recall that mutual information $I(S; X)$ is the expectation of a local function which measures the pointwise change in surprisal $i(s;x) = \Delta_s h(x)$ of value $s$ when value $x$ is observed.
Although mutual information itself is always non-negative, the pointwise function can take both positive and negative values.
Positive values correspond to a reduction in the surprisal of $s$ when $x$ is observed, negative values to an increase in surprisal.
Negative local information values are sometimes referred to as \emph{misinformation} \parencite{wibral_bits_2014} and can be interpreted as representing synergistic entropy between $S$ and $X$ \parencite{ince_partial_2017}.
Mutual information is then the expectation of both positive (information) terms and negative (misinformation) terms.
Table \ref{tab:intsign} shows how the possibility of local misinformation terms complicates pointwise interpretation of the negative local interaction information (co-information).

\begin{table}[htbp]
\begin{center}
    \begin{tabular}{| c | c | c | c | c |}
        \hline
        $\bm{\Delta_s h(x)}$ & $\bm{\Delta_s h(y)}$ & $\bm{\Delta_s h(x,y)}$ & $\bm{-i(x;y;s)}$ & \textbf{Interpretation} \\
        \hline \hline
        $+$ & $+$ & $+$ & $+$ & redundant information \\ \hline
        $+$ & $+$ & $+$ & $-$ & synergistic information \\ \hline
        $-$ & $-$ & $-$ & $-$ & redundant misinformation \\ \hline
        $-$ & $-$ & $-$ & $+$ & synergistic misinformation \\ \hline
        $+$ & $+$ & $-$ & $\dots$ & ? \\ \hline
        $-$ & $-$ & $+$ & $\dots$ & ? \\ \hline
        $+/-$ & $-/+$ & $\dots$ & $\dots$ & ? \\ \hline
  \end{tabular}
  \caption{\emph{Different interpretations of local interaction information terms.}}
  \label{tab:intsign}
\end{center}
\end{table}

Note that the fourth column represents the local co-information which quantifies the set-theoretic overlap of the two univariate local information values. 
By considering the signs of all four terms, the two univariate local informations, the local joint information and their overlap, we can determine terms which correspond to redundancy and terms which correspond to synergy.
We make an assumption that a decrease in surprisal of $s$ (positive local information term) is a fundamentally different event to an increase in surprisal of $s$ (negative local information). 
Therefore, we can only interpret the local co-information as a set-theoretic overlap in the case where all three local information terms have the same sign. 
If the joint information has a different sign to the individual informations (rows 5 and 6) the two variables together represent a fundamentally different change in surprisal than either do alone.
While a full interpretation of what these terms might represent is difficult, we argue it is clear they cannot represent a common change in surprisal. 
Similarly, if the two univariate local informations have opposite sign, they cannot have any common overlap.

The table shows that interaction information combines redundant information with synergistic misinformation, and redundant misinformation with synergistic information.
As discussed, it also includes terms which do not admit a clear interpretation.
We argue that a principled measure of redundancy should consider only redundant information and redundant misinformation.
We therefore consider the pointwise negative interaction information (overlap in surprisal), but only for symbols corresponding to the first and third rows of Table \ref{tab:intsign}.
That is, terms where the sign of the change in surprisal for all the considered sources is equal, and equal also to the sign of overlap (measured with local co-information).
In this way, we count the contributions to the overall mutual information (both positive and negative) which are genuinely shared between the input sources, while ignoring other (synergistic and ambiguous) interaction effects.
We assert that conceptually this is exactly what a redundancy function should measure.

We denote the local co-information (negative interaction information if $n$ is odd) with respect to a joint distribution $Q$ as $c_q(a_1,\dots,a_n)$, which is defined as \parencite{matsuda_physical_2000}:
\begin{align}
    c_q(a_1,\dots,a_n) = \sum_{k=1}^n (-1)^{k+1} \sum_{i_1 < \dots < i_k} h_q\left(a_{i_1},\dots,a_{i_k}\right)
    \label{eq:coinfo}
\end{align}
where $h_q(a_1,\dots,a_n)=-\log q(a_1,\dots,a_n)$ is pointwise entropy (surprisal). 
Then we define $I_\text{ccs}$, the common change in surprisal, as:
\begin{defn}

\begin{align}
    \begin{split}
        I_\text{ccs}(S;A_1,\dots,A_n) &= \sum_{a_1,\dots,a_n}\tilde{p}(a_1,\dots,a_n) \Delta_s h^\text{com}(a_1,\dots,a_n) \\
    \Delta_s h^\text{com}(a_1,\dots,a_n) &= \left\{
    \begin{array}{ll}
        c_{\tilde{p}}(a_1,\dots,a_n,s) & \begin{array}{ll} 
            \mbox{if }& \sgn \Delta_s h(a_1) = \ldots = \sgn \Delta_s h(a_n) \\
            &= \sgn \Delta_s h(a_1,\dots,a_n) =  \sgn c(a_1,\dots,a_n,s) 
        \end{array} \\
        0 & \mbox{otherwise}
    \end{array} \right.
\end{split}
\label{eq:deficcs}
\end{align}
\end{defn}
where $\Delta_s h^\text{com}(a_1,\dots,a_n)$ represents the common change in surprisal (which can be positive or negative) between the input source values, and $\tilde{P}$ is a joint distribution obtained from the observed joint distribution $P$ (see below).
$I_\text{ccs}$ measures overlapping information content with co-information, by separating contributions which correspond unambiguously to redundant mutual information at the pointwise level, and taking the expectation over these local redundancy values. 

Unlike $I_\text{min}$ which considered each input source individually, the pointwise overlap computed with local co-information  requires a joint distribution over the input sources, $\tilde{P}$ in order to obtain the local surprisal values $h_{\tilde{p}}(a_1,\dots,a_n,s)$.
We use the maximum entropy distribution subject to the constraints of equal bivariate source-target marginals, together with the equality of the n-variate joint target marginal distribution:
\begin{defn}
\begin{align}
    \begin{split}
    \hat{P}(A_1,\dots,A_n,S) &= \argmax_{Q \in \Delta_P} \sum_{a_1,\dots,a_n,s} - q(a_1,\dots,a_n,s) \log q(a_1,\dots,a_n,s) \\
    \Delta_P &= \left\{ Q \in \Delta : 
    \begin{array}{ll}
        Q(A_i,S) = P(A_i,S) \mbox{ for } i=1,\dots,n\\
        Q(A_1,\dots,A_n) = P(A_1,\dots,A_n)
    \end{array} \right\}
    \label{eq:popme}
\end{split}
\end{align}
\end{defn}
where $P(A_1,\dots,A_n,S)$ is the probability distribution defining the system under study and here $\Delta$ is the set of all possible joint distributions on $A_1,\dots,A_n,S$.
We develop the motivation for the constraints in Section~\ref{sec:gametheoretic}, and for using the distribution with maximum entropy subject to these constraints in in Section~\ref{sec:maxentsol}. 

In a previous version of this manuscript we used constraints obtained from the decision theoretic operational definition of unique information \textcite{bertschinger_quantifying_2014}.
We used the maximum entropy distribution subject to the constraints of pairwise target-predictor marginal equality:
\begin{defn}
    \label{def:pind}
\begin{align}
    \begin{split}
        \hat{P}_\text{ind}(A_1,\dots,A_n,S) &= \argmax_{Q \in \Delta_P} \sum_{a_1,\dots,a_n,s} - q(a_1,\dots,a_n,s) \log q(a_1,\dots,a_n,s) \\
    \Delta_P &= \left\{ Q \in \Delta : 
    \begin{array}{ll}
        Q(A_i,S) = P(A_i,S) \mbox{ for } i=1,\dots,n
    \end{array} \right\}
    \label{eq:pind}
\end{split}
\end{align}
\end{defn}
This illustrates $I_\text{ccs}$ can be defined in a way compatible with either operational perspective, depending on whether it is calculated using $\hat{P}$ or $\hat{P}_\text{ind}$.
We suggest that if a reader favours the decision theoretic definition of unique information \parencite{bertschinger_quantifying_2014} over the new game-theoretic definition proposed here (Section~\ref{sec:gametheoretic}) $I_\text{ccs}$ can be defined in a way consistent with that, and still provides advantages over $I_\text{broja}$, which maximises co-information without separating redundant from synergistic contributions (Section~\ref{sec:ibroja}, \ref{sec:maxentsol}).
We include Definition \ref{def:pind} here for continuity with the earlier version of this manuscript, but note that for all the examples considered here we use $\hat{P}$, following the game theoretic operational definition of unique information (Section~\ref{sec:gametheoretic}).

Note that the definition of $I_\text{min}$ in terms of minimum specific information \parencite{butts_how_2003} (Eq.~\ref{eq:imin}) suggests as a possible extension the use of a form of \emph{specific co-information}.
In order to separate redundant from synergistic components this should be thresholded with zero to only count positive (redundant) contributions.
This can be defined both in terms of target-specific co-information following $I_\text{min}$ (for clarity these definitions are shown only for two variable inputs):
\begin{equation}
    I_\text{target specific coI}(S; \mathbf{A_1}, \mathbf{A_2}) = \sum_s p(s) \max \left[ I(S=s; \mathbf{A_1}) + I(S=s; \mathbf{A_2}) - I(S=s; \mathbf{A_1}, \mathbf{A_2}), 0 \right]
\end{equation}
or alternatively in terms of source-specific co-information:
\begin{align}
    I_\text{source specific coI}(S; \mathbf{A_1}, \mathbf{A_2}) = &\sum_{a_1,a_2} p(a_1,a_2) \max \left[ I(S; \mathbf{A_1}=a_1) + I(S; \mathbf{A_2}=a_2) \right. \\
    &- \left. I(S; \mathbf{A_1}=a_1, \mathbf{A_2}=a_2), 0 \right]
\end{align}
$I_\text{ccs}$ can be seen as a fully local approach within this family of measures. 
The first key ingredient of this family is to exploit the additivity of surprisal and hence use the co-information to quantify the overlapping information content (Figure~\ref{fig:venn}); the second ingredient is to break down in some way the expectation summation in the calculation of co-information, to separate redundant and synergistic effects that are otherwise conflated.
We argue the fully local view of $I_\text{ccs}$ is required to fully separate redundant from synergistic effects. 
In either specific co-information calculation, when summing the contributions within the expectation over the non-specific variable any combination of terms listed in Table~\ref{tab:intsign} could occur.
Therefore, these specific co-information values could still conflate redundant information with synergistic misinformation.

\subsection{Calculating $I_\text{ccs}$}
\label{sec:calculationexamples}

We provide here worked examples of calculating $I_\text{ccs}$ for two simple example systems.
The simplest example of redundancy is when the system consists of a single coupled bit \parencite[Example \textsc{rdn}]{griffith_quantifying_2014}, defined by the following distribution $P(X_1,X_2,S)$:
\begin{equation}
    p(0,0,0) = p(1,1,1) = 0.5
\end{equation}
In this example $\hat{P}=P$; the maximum entropy optimisation results in the original distribution. 
Table~\ref{tab:pointwiserdn} shows the pointwise terms of the co-information calculation.
In this system for both possible configurations the change in surprisal from each predictor is $1$ bit and overlaps completely.
The signs of all changes in surprisal and the local co-information are positive, indicating that both these events correspond to redundant local information.
In this case $I_\text{ccs}$ is equal to the co-information.

\begin{table}[htbp]
\begin{center}
    \begin{tabular}{| c  | c | c | c | c | c |}
        \hline
        $\bm{(x_1,x_2,s)}$ &
        $\bm{\Delta_s h(x_1)}$ & 
        $\bm{\Delta_s h(x_2)}$ & 
        $\bm{\Delta_s h(x_1,x_2)}$ & 
        $\bm{c(x_1;x_2;s)}$ &
        $\bm{\Delta_s h^\text{com}(x_1,x_2)}$ \\
        \hline \hline
        $(0,0,0)$ & $1$  & $1$  & $1$  & $1$  & $1$ \\
        $(1,1,1)$ & $1$      & $1$      & $1$  & $1$  & $1$ \\
        \hline
  \end{tabular}
  \caption{\emph{Pointwise values from $I_\text{ccs}(S;\{1\}\{2\})$ for \textsc{rdn}}}
  \label{tab:pointwiserdn}
\end{center}
\end{table}

\begin{table}[htbp]
\begin{center}
    \begin{tabular}{| c  | c | c | c | c | c |}
        \hline
        $\bm{(x_1,x_2,s)}$ &
        $\bm{\Delta_s h(x_1)}$ & 
        $\bm{\Delta_s h(x_2)}$ & 
        $\bm{\Delta_s h(x_1,x_2)}$ & 
        $\bm{c(x_1;x_2;s)}$ &
        $\bm{\Delta_s h^\text{com}(x_1,x_2)}$ \\
        \hline \hline
        $(0,0,0)$ & $1$ & $1$ & $2$ & $0$ & $0$ \\
        $(0,1,1)$ & $0$ & $0$ & $1$ & $-1$ & $0$ \\
        $(1,0,1)$ & $0$ & $0$ & $1$ & $-1$ & $0$ \\
        $(1,1,2)$ & $1$ & $1$ & $2$ & $0$ & $0$ \\
        \hline
  \end{tabular}
  \caption{\emph{Pointwise values from $I_\text{ccs}(S;\{1\}\{2\})$ for \textsc{sum}}}
  \label{tab:pointwisesum}
\end{center}
\end{table}

The second example we consider is binary addition (see also Section~\ref{sec:sum}), $S=X_1 + X_2$, with distribution $P(X_1,X_2,S)$ given by
\begin{equation}
    p(0,0,0) = p(0,1,1) = p(1,0,1) = p(1,1,2) = \nicefrac{1}{4}
\end{equation}
In this example, again $\hat{P}=P$. 
The pointwise terms are shown in Table~\ref{tab:pointwisesum}.
For the events with $x_1=x_2$, both predictors provide $1$ bit local change in surprisal of $s$, but they do so independently since the change in surprisal when observing both together is $2$ bits.
Therefore, the local co-information is $0$; there is no overlap.
For the terms where $x_1\neq x_2$, neither predictor alone provides any local information about $s$.
However, together they provide a $1$ bit change in surprisal.
This is therefore a purely synergistic contribution, providing $-1$ bits of local co-information.
However, since this is synergistic, it is not included in $\Delta_s h^\text{com}$.
$I_\text{ccs}(S;\{1\}\{2\}) = 0$, although the co-information for this system is $-0.5$ bits. 
This example illustrates how interpreting the pointwise co-information terms allows us to select only those representing redundancy.

\subsection{Operational motivation for choice of joint distribution}
\label{sec:operational}

\subsubsection{A game-theoretic operational definition of unique information}
\label{sec:gametheoretic}

\textcite{bertschinger_quantifying_2014} introduce an operational interpretation of unique information based on decision theory, and use that to argue the ``unique and shared information should only depend on the marginal [source-target] distributions'' $P(\mathbf{A_i},S)$ (their Assumption (*) and Lemma 2).
Under the assumption that those marginals alone should specify redundancy they find $I_\text{broja}$ via maximisation of co-information.
Here we review and extend their operational argument and arrive at a different conclusion.

\textcite{bertschinger_quantifying_2014} operationalise unique information based on the idea that if an agent, Alice, has access to unique information that is not available to a second agent, Bob, there should be some situations in which Alice can exploit this information to gain a systematic advantage over Bob \parencite[Appendix B]{wibral_partial_2017}.
They formalise this as a decision problem, with the systematic advantage corresponding to a higher expected reward for Alice than Bob.
They define a decision problem as a tuple $(p,\mathcal{A},u)$ where $p(S)$ is the marginal distribution of the target, $S$, $\mathcal{A}$ is a set of possible actions the agent can take, and $u(s,a)$ is the reward function specifying the reward for each $s \in S$ $a \in \mathcal{A}$.
They assert that unique information exists if and only if there exists a decision problem in which there is higher expected reward for an agent making optimal decisions based on observation of $X_1$, versus an agent making optimal decisions on observations of $X_2$. 
This motivates their fundamental assumption that unique information depends only on the pairwise target-predictor marginals $P(X_1, S)$, $P(X_2, S)$ \parencite[Assumption (*)]{bertschinger_quantifying_2014}, and their assertion that $P(X_1,S) = P(X_2,S)$ implies no unique information in either predictor.

We argue that the decision problem they consider is too restrictive, and therefore the conclusions they draw about the properties of unique and redundant information are incorrect.
Those properties come directly from the structure of the decision problem; the reward function $u$ is the same for both agents, and the agents play independently from one other. 
The expected reward is calculated separately for each agent, ignoring by design any trial by trial covariation in their observed evidence $P(X_1,X_2)$, and resulting actions. 

While it is certainly true that if their decision problem criterion is met, then there is unique information, we argue that the decision problem advantage is not a necessary condition for the existence of unique information.
We prove this by presenting below a counter-example, in which we demonstrate unique information without a decision theoretic advantage. 
To construct this example, we extend their argument to a game-theoretic setting, where we explicitly consider two agents playing against each other. 
Decision theory is usually defined as the study of individual agents, while situations with multiple interacting agents are the purview of game theory.
Since the unique information setup includes two agents, it seems more natural to use a game theoretic approach.
Apart from switching from a decision theoretic to a game theoretic perspective, we make exactly the same argument.
It is possible to operationalise unique information so that unique information exists if and only if there exists a game (with certain properties described below) where one agent obtains a higher expected reward when both agents are playing optimally under the same utility function.

We consider two agents interacting in a game, specifically a non-cooperative, simultaneous, one-shot game \parencite{osborne_course_1994} where both agents have the same utility function. 
Non-cooperative means the players cannot form alliances or agreements.
Simultaneous (as opposed to sequential) means the players move simultaneously; if not actually simultaneous in implementation such games can be effectively simultaneous as long as each player is not aware of the other players actions.
This is a crucial requirement for a setup to operationalise unique information because if the game was sequential, it would be possible for information to `leak' from the first players evidence, via the first players action, to the second.
Restricting to simultaneous games prevents this, and ensures each game provides a fair test for unique information in each players individual predictor evidence.
One-shot (as opposed to repeated) means the game is played only once as a one off, or at least each play is completely independent of any other.
Players have no knowledge of previous iterations, or opportunity to learn from or adapt to the actions of the other player. 
The fact that the utility function is the same for the actions of each player makes it a fair test for any advantage given by unique information --- both players are playing by the same rules.
These requirements ensure that, as for the decision theoretic argument of \parencite{bertschinger_quantifying_2014}, each player must chose an action to maximise their reward based only the evidence they observe from the predictor variable.
If a player is able to obtain a systematic advantage, in the form of a higher expected reward for some specific game, given the game is fair and they are acting only on the information in the predictor they observe, then this must correspond to unique information in that predictor.
This is the same as the claim made in \parencite{bertschinger_quantifying_2014} that higher expected reward in a specific decision problem implies unique information in the predictor.  

In fact, if in addition to the above properties the considered game is also symmetric and non-zero-sum then this is exactly equivalent to the decision theoretic formulation.
Symmetric means the utility function is invariant to changes of player identity (i.e. it is the same if the players swap places). 
Alternatively, an asymmetric game is one in which the reward is not necessarily unchanged if the identity of the players is switched.
A zero-sum game is one in which there is a fixed reward that is distributed between the players\footnote{%
Technically this is a constant-sum game, but since any constant-sum game can be normalised to have zero-sum, zero-sum is the more frequently used term}, while in a non-zero-sum game the reward is not fixed.
The decision problem setup is non-zero-sum, since the action of one agent does not affect the reward obtained by the other agent.
Both players consider the game as a decision problem and so play as they would in the decision theoretic framework (i.e. to choose an action based only on their observed evidence in such a way as to maximise their expected reward). 
This is because since the game is non-cooperative, simultaneous and one-shot they have no knowledge of or exposure to the other players actions. 

We argue unique information should also be operationalised in asymmetric and zero-sum games, since these also satisfy the core requirements outlined above for a fair test of unique information. 
In a zero-sum game, the reward of each agent now also depends on the action of the other agent, therefore unique information is not invariant to changes in $P(X_1,X_2)$, because this can change the balance of rewards on individual realisations.
Note that this does not require either player is aware of the others actions (because the game is simultaneous), they still chose an action based only on their own predictor evidence, but their reward depends also on the action of the other agent (although those actions themselves are invisible).
The stochastic nature of the reward from the perspective of each individual agent is not an issue since, as for the decision theoretic approach, we consider only one-shot games. 
Alternatively, if an asymmetry is introduced to the game, for example by allowing one agent to set the stake in a gambling task, then again $P(X_1,X_2)$ affects the unique information.
We provide a specific example for this second case, and specify an actual game which meets the above requirements and provides a systematic advantage to one player, demonstrating the presence of unique information.
However, this system does not admit a decision problem which provides an advantage.
This counter-example therefore proves that the decision theoretic operationalisation of \parencite{bertschinger_quantifying_2014} is not a necessary condition for the existence of unique information.

Borrowing notation from \parencite{bertschinger_quantifying_2014} we consider two agents, which each observe values from $X_1$ and $X_2$ respectively, and take actions $a_1,a_2 \in \mathcal{A}$. 
Both are subject the same core utility function $v(s,a)$, but we break the symmetry in the game by allowing one agent to perform a second action - setting the stake on each hand (realisation).
This results in utility functions $u_i(s,a_i,x_1) = c(x_1) v(s, a_i)$, where $c$ is a stake weighting chosen by agent 1 on the basis of their evidence. 
This stake weighting is not related to their guess on the value $s$ (their action $a_i$), but serves here as a way to break the symmetry of the game while maintaining equal utility functions for each player.
That is, although the reward here is a function also of $x_1$, it is the same function for both players, so $a_1=a_2 \implies u_1(s,a_1,x_1)=u_2(s,a_2,x_1) \forall s,x_1$.
In general, in the game theoretic setting the utility function can depend on the entire state of the world, $u(s,a_i,x_1,x_2)$, but here we introduce only an asymmetric dependence on $x_1$. 
Both agents have the same utility function as required for a fair test of unique information, but that utility function is asymmetric --- it is not invariant to switching the players.
The second agent is not aware of the stake weighting applied to the game when they choose their action.
The tuple $(p,\mathcal{A},\mathbf{u})$ defines the game with $\mathbf{u}(s,a_1,a_2,x_1) = [ u_1(s,a_1,x_1),  u_2(s,a_2,x_1) ]$.
In this case the reward of agent 2 depends on $x_1$, introducing again a dependence on $P(X_1,X_,2)$.
However, because both agents have the same asymmetric utility function, this game meets the intuitive requirements for an operational test of unique information. 
If there is no unique information, agent 1 should not be able to profit simply by changing the stakes on different trials. 
If they can profit systematically by changing the stakes on trials that are favourable to them based on the evidence they observe, that is surely an operationalisation of unique information.
We emphasise again that we are considering here a non-cooperative, simultaneous, one-shot, non-zero-sum, asymmetric game.
So agent 2 does not have any information about the stake weight on individual games, and cannot learn anything about the stake weight from repeated plays. 
Therefore, there is no way for unique information in $X_1$ to affect the action of agent 2 via the stake weight setting.
The only difference from the decision theoretic framework is that here we consider an asymmetric utility function.
 
\begin{figure}[htbp]
    \centering
    \includegraphics[width=0.5\textwidth]{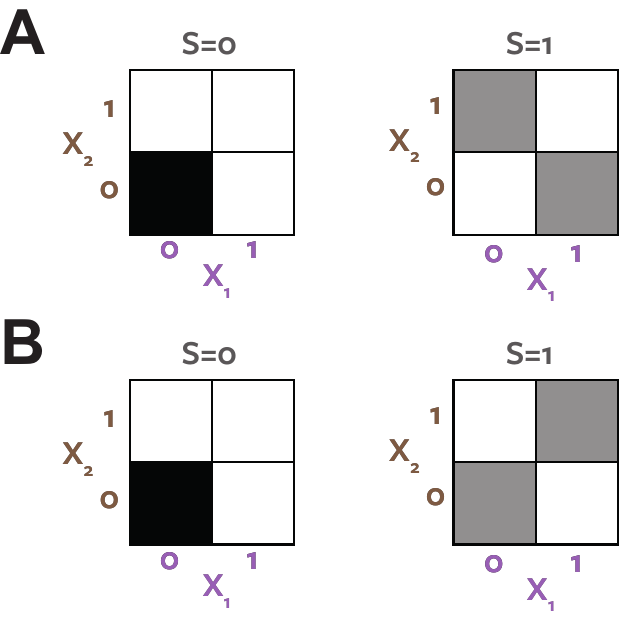}
    \caption{\emph{\textsc{ReducedOr}}. 
    \textbf{A} Probability distribution of \textsc{ReducedOr} system.
    \textbf{B} Distribution resulting from $I_\text{broja}$ optimisation. 
    Black tiles represent outcomes with $p=0.5$.
    Grey tiles represent outcomes with $p=0.25$.
    White tiles are zero-probability outcomes.}
    \label{fig:reducedor}
\end{figure}

To demonstrate this, and provide a concrete counter-example to the decision theoretic argument \parencite{bertschinger_quantifying_2014} we consider a system termed \textsc{ReducedOr}\footnote{Joseph Lizier, \emph{personal communication}}.
Figure~\ref{fig:reducedor}A shows the probability distribution which defines this binary system. 
Figure~\ref{fig:reducedor}B shows the distribution resulting from the $I_\text{broja}$ optimisation procedure. 
Both systems have the same target-predictor marginals $P(X_i,S)$, but have different predictor-predictor marginals $P(X_1,X_2)$. 
$I_\text{broja}$ reports zero unique information.
$I_\text{ccs}$ reports zero redundancy, but unique information present in both predictors.

\begin{table}[htbp]
\begin{center}
    \begin{tabular}{| c  | c | c | c |}
        \hline
        \textbf{Node} &  ${I_\partial} [ I_\text{min} ]$ & ${I_\partial} [ I_\text{broja} ]$ & ${I_\partial} [ I_\text{ccs} ]$ \\
        \hline \hline
        $\{1\}\{2\}$ & $0.31$ & $0.31$ & $0$\\
        $\{1\}$ & $0$ & $0$ & $0.31$ \\
        $\{2\}$ & $0$ & $0$ & $0.31$ \\ 
        $\{12\}$ & $0.69$ & $0.69$ & $0.38$ \\ \hline
  \end{tabular}
  \caption{\emph{PIDs for \textsc{ReducedOr} (Figure~\ref{fig:reducedor}A)}}
  \label{tab:reducedor}
\end{center}
\end{table}

In the $I_\text{broja}$ optimised distribution (Figure~\ref{fig:reducedor}B) the two predictors are directly coupled, $P(X_1=0,X_2=1) = P(X_1=1,X_2=0) = 0$. 
In this case there is clearly no unique information. 
The coupled marginals mean both agents see the same evidence on each realisation, make the same choice and therefore obtain the same reward, regardless of the stake weighting chosen by agent 1. 
However, in the actual system, the situation is different. 
Now the evidence is de-coupled, the agents never both see the evidence $x_i=1$ on any particular realisation $P(X_1=1,X_2=1)=0$.
Assuming a utility function $v(s,a)=\delta_{sa}$ reflecting a guessing game task, the optimal strategy for both agents is to make a guess $a_i=0$ when they observe $x_i=0$, and guess $a_i=1$ when they observe $x_i=1$.
If Alice ($X_1$) controls the stake weight she can choose $c(x_1)=1+x_1$ which results in a doubling of the reward when she observes $X_1=1$ versus when she observes $X_1=0$.
Under the true distribution of the system for realisations where $x_1=1$, we know that $x_2=0$ and $s=1$, so Bob will guess $a_2=0$ and be wrong (have zero reward).
On an equal number of trials Bob will see $x_2=1$, guess correctly and Alice will win nothing, but those trials have half the utility of the trials that Alice wins due to the asymmetry resulting from her specifying the gambling stake. 
Therefore, on average, Alice will have a systematically higher reward as a result of exploiting her unique information, which is unique because on specific realisations it is available only to her.
Similarly, the argument can be reversed, and if Bob gets to choose the stakes, corresponding to a utility weighting $c(x_2)=1+x_2$, he can exploit unique information available to him on a separate set of realisations.

Both games considered above would provide no advantage when applied to the $I_\text{broja}$ distribution (Figure~\ref{fig:reducedor}B).
The information available to each agent when they observe $X_i=1$ is not unique, because it always occurs together on the same realisations.
There is no way to gain an advantage in any game since it will always be available simultaneously to the other agent.
In both decompositions the information corresponding to prediction of the stimulus when $x_i=1$ is quantified as $0.31$ bits. 
$I_\text{broja}$ quantifies this as redundancy because it ignores the structure of $P(X_1,X_2)$ and so does not consider the within trial relationships between the agents evidence.
$I_\text{broja}$ cannot distinguish between the two distributions illustrated in Figure~\ref{fig:reducedor}.
$I_\text{ccs}$ quantifies the $0.31$ bits as unique information in both predictors, because in the true system each agent sees the informative evidence on different trials, and so can exploit it to gain a higher reward in a certain game.
$I_\text{ccs}$ agrees with $I_\text{broja}$ in the system in Figure~\ref{fig:reducedor}B, because here the same evidence is always available to both agents, so is not unique.

We argue that this example directly  illustrates the fact that unique information is not invariant to $P(X_1,X_2)$, and that the decision theoretic operational definition of \parencite{bertschinger_quantifying_2014} is too restrictive. 
The decision theory view says that unique information corresponds to an advantage which can be obtained only when two players go to different private rooms in a casino, play independently and then compare their winnings at the end of the session. 
The game theoretic view says that unique information corresponds to any obtainable advantage in a fair game (simultaneous and with equal utility functions), even when the players play each other directly, betting with a fixed pot, on the same hands at the same table.
We have shown a specific example where there is an advantage in the second case, but not the first case. 
We suggest such an advantage cannot arise without unique information in the predictor and therefore claim this counter-example proves that the decision theoretic operationalisation is not a necessary condition for the existence of unique information.
While this is a single specific system, we will see in the examples (Section~\ref{sec:examples}) that the phenomenon of $I_\text{broja}$ over-stating redundancy by neglecting unique information which is masked when the inputs are coupled occurs frequently.
We argue this occurs because the $I_\text{broja}$ optimisation maximises co-information.
It therefore couples the predictors to maximise the contribution of source redundancy to the co-information, since the game theoretic operationalisation shows that redundancy is not invariant to the predictor-predictor marginal distribution. 

\subsubsection{Maximum entropy solution}
\label{sec:maxentsol}

For simplicity we consider first a two-predictor system.
The game-theoretic operational definition of unique information provided in the previous section requires that the unique information (and hence redundancy) should depend only on the pairwise marginals $P(S,X_1)$, $P(S,X_2)$ and $P(X_1,X_2)$. 
Therefore, any measure of redundancy which is consistent with this operational definition should take a constant value over the family of distributions which satisfy those marginal constraints.
This is the same argument applied in \parencite{bertschinger_quantifying_2014} but we consider here the game-theoretic extension to their decision theoretic operationalisation.
Co-information itself is not constant over this family of distributions, because its value can be altered by third order interactions (i.e. those not specified by the pairwise marginals)\footnote{%
Consider for example \textsc{xor}. The co-information of this distribution is $-1$ bits, but the maximum entropy distribution preserving pairwise marginal constraints is the uniform distribution with a co-information of $0$ bits}.
Therefore, if $I_\text{ccs}$ were calculated using the full joint distribution it would not be consistent with the game-theoretic operational definition of unique information.  

Since redundancy should be invariant given the specified marginals, our definition of $I_\text{ccs}$ must be a function only of those marginals.
However, we need a full joint distribution over the trivariate joint space to calculate the pointwise co-information terms.
We use the maximum entropy distribution subject to the constraints specified by the game-theoretic operational definition (Eq.~\ref{eq:popme}).
The maximum entropy distribution is by definition the most parsimonious way to fill out a full trivariate distribution given only a set of bi-variate marginals \parencite{jaynes_information_1957}.
It introduces no additional structure to the 3-way distribution over that which is specified by the constraints.
Pairwise marginal constrained maximum entropy distributions have been widely used to study the effect of third and higher order interactions, because they provide a surrogate model which removes these effects \parencite{amari_information_2010,schneidman_network_2003,ince_presence_2009,roudi_pairwise_2009}. 
Any distribution with lower entropy would by definition have some additional structure over that which is required to specify the unique and redundant information following the game-theoretic operationalisation. 

Note that the definition of $I_\text{broja}$ follows a similar argument. 
If redundancy was measured with co-information directly, it would not be consistent with the decision theoretic operationalisation \parencite{bertschinger_quantifying_2014}.
\textcite{bertschinger_quantifying_2014} address this by choosing the distribution which maximises co-information subject to the decision theoretic constraints.
While we argue that maximizing entropy is in general a more principled approach than maximizing co-information, note that with the additional predictor marginal constraint introduced by the game-theoretic operational definition, both approaches are equivalent for two predictors (since maximizing co-information is equal to maximizing entropy given the constraints). 
However, once the distribution is obtained the other crucial difference is that $I_\text{ccs}$ separates genuine redundant contributions at the local level, while $I_\text{broja}$ computes the full co-information, which conflates redundant and synergistic effects (Table~\ref{tab:intsign}) \parencite{williams_nonnegative_2010}.

We apply our game-theoretic operational definition in the same way to provide the constraints in Eq.~\ref{eq:popme} for an arbitrary number of inputs.
The action of each agent is determined by $P(A_i,S)$ (or equivalently $P(S|A_i)$) and the agent interaction effects (from zero-sum or asymmetric utility functions) are determined by $P(A_1,\dots,A_n)$.

\subsection{Properties}
\label{sec:properties}

The measure $I_{\text{ccs}}$ as defined above satisfies some of the proposed redundancy axioms (Section~\ref{sec:pid}).
The symmetry and self-redundancy axioms are satisfied from the properties of co-information \parencite{matsuda_physical_2000}.
For self-redundancy, consider that co-information for $n=2$ is equal to mutual information at the pointwise level (Eq.~\ref{eq:coinfo}):
\begin{align}
    \begin{split}
    c(s,a) &= h(s) + h(a) - h(s,a) \\
    &= i(s;a) = \Delta_s h(a)
\end{split}
\end{align}
So $\sgn c(s,a)=\sgn \Delta_s h(a) \; \forall s,a$ and $I_\text{ccs}(S;A) = I(S;A)$.
Subset equality is also satisfied.
If $\mathbf{A_{l-1}} \subseteq \mathbf{A_l}$ then we consider values
$a_{l-1} \in \mathbf{A_{l-1}}$, $a_{l} \in \mathbf{A_{l}}$ with $a_l = (a_l^{l-1},a_l^+)$ and $a_l^{l-1} \in \mathbf{A_{l-1}} \cap \mathbf{A_l} = \mathbf{A_{l-1}}$, $a_l^+ \in \mathbf{A_l} \setminus \mathbf{A_{l-1}}$. 
Then
\begin{align}
\begin{split}
    p(a_{i_1},\dots,a_{i_j},a_{l-1},a_l^{l-1},a_l^+) = 
    \begin{cases}
        0 &\text{if } a_{l-1} \neq a_l^{l-1} \\
        p(a_{i_1},\dots,a_{i_j},a_l) &\text{otherwise}   
\end{cases}
\end{split}
\end{align}
for any $i_1<\dots<i_j \in \{1,\dots,l-2\}$.
So for non-zero terms in Eq.~\ref{eq:deficcs}:
\begin{equation}
h(a_{i_1},\dots,a_{i_j},a_{l-1},a_l) = h(a_{i_1},\dots,a_{i_j},a_l) 
\end{equation}
 
Therefore all terms for $k \geq 2$ in Eq.~\ref{eq:coinfo} which include $a_{l-1},a_l$ cancel with a corresponding $k-1$ order term including $a_l$, so
\begin{equation}
    c(a_1,\dots,a_{l-1},a_l) = c(a_1,\dots,a_{l-1})
\end{equation}
and subset equality holds.

$I_\text{ccs}$ does not satisfy the Harder et al.~identity axiom \textcite{harder_bivariate_2013} (Eq.~\ref{eq:harderidentity}); any distribution with negative local information terms serves as a counter example.
These negative terms represent synergistic entropy which is included the standard mutual information quantity \parencite{ince_partial_2017}.
Therefore their omission in the calculation of $I_\text{ccs}$ seems appealing; since they result from a synergistic interaction they should not be included in a measure quantifying redundant information. 
$I_\text{ccs}$ does satisfy the modified independent identity axiom (Eq.~\ref{eq:indidentity}), and so correctly quantifies redundancy in the two-bit copy problem (Section~\ref{sec:imin}).

However, $I_{\text{ccs}}$ does not satisfy monotonicity.
To demonstrate this, consider the following example \parencite[Table \ref{tab:uniquemisex}, modified from][Figure 3]{griffith_intersection_2014}.
\begin{table}[htbp]
\begin{center}
    \begin{tabular}{| c  c  c || c |}
        \hline
        $x_1$ & $x_2$ & $s$ & $p(x_1, x_2, s)$ \\
        \hline \hline
        $0$ & $0$ & $0$ & $0.4$ \\
        $0$ & $1$ & $0$ & $0.1$ \\
        $1$ & $1$ & $1$ & $0.5$ \\ \hline
  \end{tabular}
  \caption{\emph{Example system with unique misinformation.}}
  \label{tab:uniquemisex}
\end{center}
\end{table}

For this system, 
\begin{align*}
    I(S; X_1) = I(S; X_1, X_2) &= 1 \mbox{ bit}\\
    I(S;X_2) = 0.61 \mbox{ bits}
\end{align*}
Because of the self redundancy property, these values specify $I_\cap$ for the upper 3 values of the redundancy lattice (Figure \ref{fig:lattice}A).
The value of the bottom node is given by 
\begin{align*}
    I_\partial = I_\cap = I_\text{ccs}(S; \{1\}\{2\}) = 0.77 \mbox{ bits}
\end{align*}
This value arises from two positive pointwise terms:
\begin{align*}
    x_1=x_2=s&=0 \mbox{ (contributes $0.4$ bits)} \\
    x_1=x_2=s&=1 \mbox{ (contributes $0.37$ bits)}
\end{align*}
So $I_\text{ccs}(S; \{1\}\{2\}) > I_\text{ccs}(S;\{2\})$ which violates monotonicity on the lattice.
How is it possible for two variables to share more information than one of them carries alone?

Consider the pointwise mutual information values for $I_\text{ccs}(S; \{2\}) = I(S;X_2)$.
There are the same two positive information terms that contribute to the redundancy (since both are common with $X_1$).
However, there is also a third misinformation term of $-0.16$ bits when $s=0, x_2=1$.
In our view, this demonstrates that the monotonicity axiom is incorrect for a measure of redundant information content.
As this example shows a node can have \emph{unique misinformation}.

For this example $I_\text{ccs}$ yields the PID: 
\begin{align*}
    I_\partial(\{1\}\{2\}) &= 0.77 \\
    I_\partial(\{1\}) &= 0.23 \\
    I_\partial(\{2\}) &= -0.16 \\
    I_\partial(\{12\}) &= 0.16
\end{align*}

While monotonicity has been considered a crucial axiom with the PID framework, we argue that subset equality, usually considered as part of the axiom of monotonicity, is the essential property that permits the use of the redundancy lattice.
We have seen this lack of monotonicity means the PID obtained with $I_\text{ccs}$ is not non-negative. 
We agree  that while ``negative \dots~atoms can \emph{subjectively} be seen as flaw'' \parencite{james_multivariate_2016}, we argue here that in fact they are a necessary consequence of a redundancy measure that genuinely quantifies overlapping information content\footnote{%
Please note that in an earlier version of this manuscript we proposed thresholding with 0 to remove negative values. We no longer do so.}.
Mutual information is the expectation of a local quantity that can take both positive (local information) and negative (local misinformation) values, corresponding to redundant and synergistic entropy respectively \parencite{ince_partial_2017}.
Jensen's equality ensures that the final expectation value of mutual information is positive; or equivalently that redundant entropy is greater than synergistic entropy in any bivariate system.
We argue that when breaking down the classical Shannon information into a partial information decomposition, there is no reason that those partial information values must be non-negative, since there is no way to apply Jensen's inequality to these partial values. 
We have illustrated this with a simple example where a negative unique information value is obtained, and inspection of the pointwise terms shows that this is indeed due to negative pointwise terms in the mutual information calculation for one predictor that are not present in the mutual information calculation for the other predictor: unique misinformation. 
Applying the redundancy lattice and the partial information decomposition directly to entropy can provide some further insights into the prevalence and effects of misinformation or synergistic entropy \parencite{ince_partial_2017}. 

We conjecture that $I_\text{ccs}$ is continuous in the underlying probability distribution \parencite{lizier_towards_2013-1} from the continuity of the logarithm and co-information, but not differentiable due to the thresholding with $0$.
Continuity requires that, at the local level, 
\begin{equation}
    c(s,a_1,a_2) < \min \left[ i(s;a_1), i(s;a_2) \right]
    \label{eq:overlaplarger}
\end{equation}
when $\sgn i(s;a_1) =  \sgn i(s; a_2) = \sgn i(s; a_1,a_2) = \sgn c(s,a_1,a_2)$.
While this relationship holds for the full integrated quantities \parencite{matsuda_physical_2000}, it does not hold at the local level for all joint distributions.
However, we conjecture that it holds when using the pairwise maximum entropy solution $\hat{P}$, with no higher order interactions.  
This is equivalent to saying that the overlap of the two local informations should not be larger than the smallest --- an intuitive requirement for a set theoretic overlap. 
However, at this stage the claim of continuity remains a conjecture.
In the Matlab implementation we explicitly test for violations of the condition in Eq.~\ref{eq:overlaplarger}, which do not occur in any of the examples we consider here.
This shows that all the examples we consider here are at least locally continuous in the neighbourhood of the specific joint probability distribution considered.

In the next sections, we demonstrate with a range of example systems how the results obtained with this approach match intuitive expectations for a partial information decomposition.

\subsection{Implementation}

Matlab code is provided to accompany this article, which features simple functions for calculating the partial information decomposition for two and three variables\footnote{%
Available at: \href{https://github.com/robince/partial-info-decomp}{\texttt{https://github.com/robince/partial-info-decomp}}.
}.
This includes implementation of $I_\text{min}$ and the PID calculation of \textcite{williams_nonnegative_2010}, as well as $I_\text{ccs}$ and $I_\text{broja}$. 
Scripts are provided reproducing all the examples considered here.
Implementations of $I_\text{ccs}$ and $I_\text{mmi}$ \parencite{barrett_exploration_2015} for Gaussian systems are also included.

To calculate $I_\text{broja}$ and compute the maximum entropy distributions under marginal constraints we use the \texttt{dit} package \parencite{james_dit/dit_2017}\footnote{\href{https://github.com/dit/dit}{\texttt{https://github.com/dit/dit}}
\hspace{0.2cm}  \href{http://docs.dit.io/}{\texttt{http://docs.dit.io/}}}.

\section{Two variable examples}
\label{sec:examples}

\subsection{Examples from \textcite{williams_nonnegative_2010}}

\begin{figure}[htbp]
    \centering
    \includegraphics[width=0.5\textwidth]{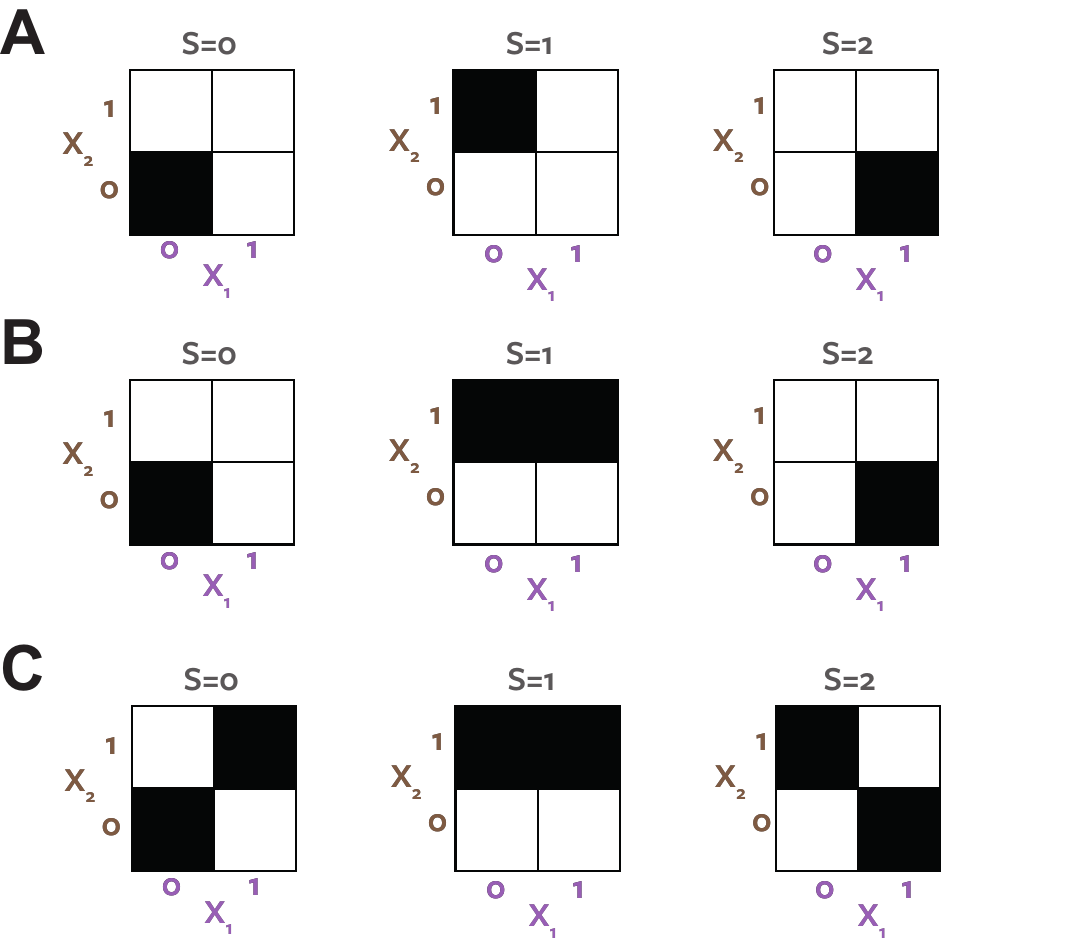}
    \caption{\emph{Probability distributions for three example systems}. 
    Black tiles represent equiprobable outcomes.
    White tiles are zero-probability outcomes.
    A and B modified from \textcite{williams_nonnegative_2010}.}
    \label{fig:wbexamples}
\end{figure}

We begin with the original examples of \textcite[][Figure 4]{williams_nonnegative_2010}, reproduced here in Figure \ref{fig:wbexamples}.

\begin{table}[htbp]
\begin{center}
    \begin{tabular}{| c  | c | c | c |}
        \hline
        \textbf{Node} & ${I_\partial} [ I_\text{min} ]$ & ${I_\partial} [ I_\text{broja} ]$ & ${I_\partial} [ I_\text{ccs} ]$ \\
        \hline \hline
        $\{1\}\{2\}$  & $0.5850$                        & $0.2516$                          & $0.3900$                        \\
        $\{1\}$       & $0.3333$                        & $0.6667$                          & $0.5283$                        \\
        $\{2\}$       & $0.3333$                        & $0.6667$                          & $0.5283$                        \\
        $\{12\}$      & $0.3333$                        & $0$                               & $0.1383$                        \\ \hline
  \end{tabular}
  \caption{\emph{PIDs for example Figure \ref{fig:wbexamples}A}}
  \label{tab:wbA}
\end{center}
\end{table}

Table \ref{tab:wbA} shows the PIDs for the system shown in \ref{fig:wbexamples}A, obtained with $I_\text{min}$, $I_\text{broja}$ and $I_\text{ccs}$\footnote{%
This is equivalent to the system \textsc{Subtle} in \textcite[][Figure 4]{griffith_intersection_2014}.}.
$I_\text{ccs}$ and $I_\text{min}$ agree qualitatively here; both show both synergistic and redundant information.
$I_\text{broja}$ shows zero synergy.
The pointwise computation of $I_\text{ccs}$ includes two non-zero terms; when 
\begin{align*}
    x_1=0,x_2=1,s&=1 \mbox{ and when}  \\
    x_1=1,x_2=0,s&=2
\end{align*}
For both of these local values, $x_1$ and $x_2$ are contributing the same reduction in surprisal of $s$ ($0.195$ bits each for $0.39$ bits overall redundancy).
There are no other redundant local changes in surprisal (positive or negative).
In this case, both the $I_\text{broja}$ optimised distribution and the pairwise marginal maximum entropy distribution are equal to the original distribution.
So here $I_\text{broja}$ is measuring redundancy directly with co-information, whereas $I_\text{ccs}$ breaks down the co-information to include only the two terms which directly represent redundancy.
In the full co-information calculation of $I_\text{broja}$ there is one additional contribution of $-0.138$ bits, which comes from the $x_1=x_2=s=0$ event.
In this case the local changes in surprisal of $s$ from $x_1$ and $x_2$ are both positive ($0.585$), but the local co-information is negative ($-0.415$).
This corresponds to the second row of Table~\ref{tab:intsign} --- it is synergistic local information.
Therefore this example clearly shows how the $I_\text{broja}$ measure of redundancy erroneously includes synergistic effects.

Table \ref{tab:wbB} shows the PIDs for the system shown in \ref{fig:wbexamples}B.
Here $I_\text{broja}$ and $I_\text{ccs}$ agree, but diverge qualitatively from $I_\text{min}$.
$I_\text{min}$ shows both synergy and redundancy, with no unique information carried by $X_1$ alone.
$I_\text{ccs}$ shows no synergy and redundancy, only unique information carried independently by $X_1$ and $X_2$.
\textcite{williams_nonnegative_2010} argue that ``$X_1$ and $X_2$ provide 0.5 bits of redundant information corresponding to the fact that knowledge of either $X_1$ or $X_2$ reduces uncertainty about the outcomes $S=0,S=2$''.
However, while both variables reduce uncertainty about $S$, they do so in different ways --- $X_1$ discriminates the possibilities $S=0,1$ vs.~$S=1,2$ while $X_2$ allows discrimination between $S=1$ vs.~$S=0,2$.
These discriminations represent different non-overlapping information content, and therefore should be allocated as unique information to each variable as in the $I_\text{ccs}$ and $I_\text{broja}$ PIDs.
While the full outcome can only be determined with knowledge of both variables, there is no synergistic information because the discriminations described above are independent.

\begin{table}[htbp]
\begin{center}
    \begin{tabular}{| c  | c | c | c |}
        \hline
        \textbf{Node} & ${I_\partial} [ I_\text{min} ]$ & ${I_\partial} [ I_\text{broja} ]$ & ${I_\partial} [ I_\text{ccs} ]$ \\
        \hline \hline
        $\{1\}\{2\}$  & $0.5$                        & $0$                          & $0$                        \\
        $\{1\}$       & $0$                        & $0.5$                          & $0.5$                        \\
        $\{2\}$       & $0.5$                        & $1$                          & $1$                        \\
        $\{12\}$      & $0.5$                        & $0$                               & $0$                        \\ \hline
  \end{tabular}
  \caption{\emph{PIDs for example Figure \ref{fig:wbexamples}B}}
  \label{tab:wbB}
\end{center}
\end{table}

To induce genuine synergy it is necessary to make the $X_1$ discrimination between $S=0,1$ and $S=1,2$ ambiguous without knowledge of $X_2$.
Table \ref{tab:wbC} shows the PID for the system shown in \ref{fig:wbexamples}C, which includes such an ambiguity.
Now there is no information in $X_1$ alone, but it contributes synergistic information when $X_2$ is known.
Here, $I_\text{min}$ correctly measures $0$ bits redundancy, and all three PIDs agree (the other three terms have only one source, and therefore are the same for all measures from self-redundancy).

\begin{table}[htbp]
\begin{center}
    \begin{tabular}{| c  | c | c | c |}
        \hline
        \textbf{Node} & ${I_\partial} [ I_\text{min} ]$ & ${I_\partial} [ I_\text{broja} ]$ & ${I_\partial} [ I_\text{ccs} ]$ \\
        \hline \hline
        $\{1\}\{2\}$  & $0$                        & $0$                          & $0$                        \\
        $\{1\}$       & $0$                        & $0$                          & $0$                        \\
        $\{2\}$       & $0.25$                        & $0.25$                          & $0.25$                        \\
        $\{12\}$      & $0.67$                        & $0.67$                               & $0.67$                        \\ \hline
  \end{tabular}
  \caption{\emph{PIDs for example Figure \ref{fig:wbexamples}C}}
  \label{tab:wbC}
\end{center}
\end{table}

\subsection{Binary logical operators}

\begin{figure}[htbp]
    \centering
    \includegraphics[width=0.4\textwidth]{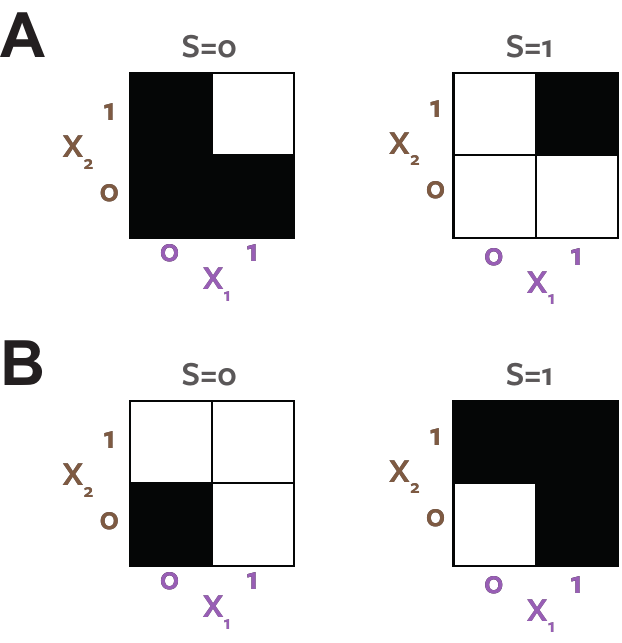}
    \caption{\emph{Binary logical operators.}
    Probability distributions for \textbf{A}: \textsc{and}, \textbf{B}: \textsc{or}. 
    Black tiles represent equiprobable outcomes.
    White tiles are zero-probability outcomes.}
    \label{fig:andor}
\end{figure}

The binary logical operators \textsc{or}, \textsc{xor} and \textsc{and} are often used as example systems \parencite{harder_bivariate_2013,griffith_quantifying_2014,bertschinger_quantifying_2014}.
For \textsc{xor}, the $I_\text{ccs}$ PID agrees with both $I_\text{min}$ and $I_\text{broja}$ and quantifies the $1$ bit of information as fully synergistic.

\subsubsection{AND/OR}
\label{sec:and}

Figure \ref{fig:andor} illustrates the probability distributions for \textsc{and} and \textsc{or}.
This makes clear the equivalence between them; because of symmetry any PID should give the same result on both systems.
Table \ref{tab:andor} shows the PIDs.

\begin{table}[htbp]
\begin{center}
    \begin{tabular}{| c  | c | c | c |}
        \hline
        \textbf{Node} & ${I_\partial} [ I_\text{min} ]$ & ${I_\partial} [ I_\text{broja} ]$ & ${I_\partial} [ I_\text{ccs} ]$ \\
        \hline \hline
        $\{1\}\{2\}$  & $0.31$                          & $0.31$                            & $0.10$                             \\
        $\{1\}$       & $0$                             & $0$                               & $0.21$                             \\
        $\{2\}$       & $0$                             & $0$                               & $0.21$                          \\
        $\{12\}$      & $0.5$                           & $0.5$                             & $0.29$                          \\ \hline
  \end{tabular}
  \caption{\emph{PIDs for \textsc{and/or}}}
  \label{tab:andor}
\end{center}
\end{table}

In this system $I_\text{min}$ and $I_\text{broja}$ agree, both showing no unique information.
$I_\text{ccs}$ shows less redundancy, and unique information in both predictors.
The redundancy value with $I_\text{ccs}$ falls within the bounds proposed in \textcite[][Figure 6.11]{griffith_quantifying_2014}.

\begin{table}[htbp]
\begin{center}
    \begin{tabular}{| c  | c | c | c | c | c |}
        \hline
        $\bm{(x_1,x_2,s)}$ &
        $\bm{\Delta_s h(x_1)}$ & 
        $\bm{\Delta_s h(x_2)}$ & 
        $\bm{\Delta_s h(x_1,x_2)}$ & 
        $\bm{c(x_1;x_2;s)}$ &
        $\bm{\Delta_s h^\text{com}(x_1,x_2)}$ \\
        \hline \hline
        $(0,0,0)$ & $0.415$  & $0.415$  & $0.415$  & $0.415$  & $0.415$ \\
        $(0,1,0)$ & $0.415$  & $-0.585$ & $0.415$  & $-0.585$ & $0$     \\
        $(1,0,0)$ & $-0.585$ & $0.415$  & $0.415$  & $-0.585$ & $0$     \\
        $(1,1,1)$ & $1$      & $1$      & $2$  & $0$  & $0$ \\
        \hline
  \end{tabular}
  \caption{\emph{Pointwise values from $I_\text{ccs}(S;\{1\}\{2\})$ for \textsc{and}}}
  \label{tab:pointwiseand}
\end{center}
\end{table}

To see where this unique information arises with $I_\text{ccs}$ we can consider directly the individual pointwise contributions for the \textsc{and} example (Table \ref{tab:pointwiseand}).
$I_\text{ccs}(\{1\}\{2\})$ has a single pointwise contribution from the event $(0,0,0)$, only when both inputs are $0$ is there redundant local information about the outcome.
For the event $(0,1,0)$ (and symmetrically for $1,0,0$) $x_1$ conveys local information about $s$, while $x_2$ conveys local misinformation, therefore there is no redundancy, but a unique contribution for both $x_1$ and $x_2$.
We can see in the $(1,1,1)$ event the change in surprisal of $s$ from the two predictors is independent, so again contributes unique rather than redundant information.
So the unique information in each predictor is a combination of unique information and misinformation terms.

For $I_\text{broja}$ the specific joint distribution that maximises the co-information in the \textsc{and} example while preserving $P(X_i,S)$  \parencite[Example 30, $\alpha=\nicefrac{1}{4}$]{bertschinger_quantifying_2014} has an entropy of $1.5$~bits.
$\hat{P}(X_1,X_2,S)$ used in the calculation of $I_\text{ccs}$ is equal to the original distribution and has an entropy of $2$ bits.
Therefore, the distribution used in $I_\text{broja}$ has some additional structure above that specified by the individual joint target marginals and which is chosen to maximise the co-information (negative interaction information).
As discussed above, interaction information can conflate redundant information with synergistic misinformation, as well as having other ambiguous terms when the signs of the individual changes of surprisal are not equal.
As shown in Table \ref{tab:pointwiseand}, the \textsc{and} system includes such ambiguous terms (rows 2 and 3, which contribute synergy to the interaction information), and also includes some synergistic misinformation (row 4, which contributes redundancy to the interaction information). 
Any system of the form considered in \textcite[][Example 30]{bertschinger_quantifying_2014} will have similar contributing terms.
This illustrates the problem with using co-information directly as a redundancy measure, regardless of how the underlying distribution is obtained.
The distribution selected to maximise co-information will be affected by these ambiguous and synergistic terms.
In fact, it is interesting to note that for the $I_\text{broja}$ distribution ($\alpha=\nicefrac{1}{4}$), $p(0,1,0)=p(1,0,0)=0$ and the two ambiguous synergistic terms are removed from the interaction information.
This indicates how the optimisation of the co-information might be driven by terms that cannot be interpreted as genuine redundancy.
Further, the distribution used in $I_\text{broja}$ has perfectly coupled marginals.
This increases the source redundancy measured by the co-information.
Under this distribution, the $(1,1,1)$ term now contributes $1$ bit locally to the co-information.
This is redundant because $x_1=1$ and $x_2=1$ always occur together.
In the original distribution the $(1,1,1)$ term is independent because the predictors are independent.

We argue there is no fundamental conceptual problem with the presence of unique information in the \textsc{and} example.
Both variables share some information, have some synergistic information, but also have some unique information corresponding to the fact that knowledge of either variable taking the value $1$ reduces the uncertainty of $s=1$ independently (i.e. on different trials).
If the joint target marginal distributions are equal, then by symmetry $I_\partial(\{1\}) = I_\partial(\{2\})$, but it is not necessary that $I_\partial(\{1\}) = I_\partial(\{2\}) = 0$ \parencite[][Corollary 8]{bertschinger_quantifying_2014}.

\subsubsection{SUM}
\label{sec:sum}

While not strictly a binary logic gate, we also consider the summation of two binary inputs. 
The \textsc{and} gate can be thought of as a thresholded version of summation.
Summation of two binary inputs is also equivalent to the system \textsc{XorAnd} \parencite{harder_bivariate_2013,griffith_quantifying_2014,bertschinger_quantifying_2014}.
Table~\ref{tab:sum} shows the PIDs.

\begin{table}[htbp]
\begin{center}
    \begin{tabular}{| c  | c | c | c |}
        \hline
        \textbf{Node} & ${I_\partial} [ I_\text{min} ]$ & ${I_\partial} [ I_\text{broja} ]$ & ${I_\partial} [ I_\text{ccs} ]$ \\
        \hline \hline
        $\{1\}\{2\}$  & $0.5$                           & $0.5$                             & $0$                             \\
        $\{1\}$       & $0$                             & $0$                               & $0.5$                           \\
        $\{2\}$       & $0$                             & $0$                               & $0.5$                           \\
        $\{12\}$      & $1$                             & $1$                               & $0.5$                           \\ \hline
  \end{tabular}
  \caption{\emph{PIDs for \textsc{sum}}.}
  \label{tab:sum}
\end{center}
\end{table}

As with \textsc{and}, $I_\text{min}$ and $I_\text{broja}$ agree, and both allocate $0$ bits of unique information.
Both of these methods always allocate zero unique information when the target-predictor marginals are equal.
$I_\text{ccs}$ differs in that it allocates $0$ redundancy.
This arises for a similar reason to the differences discussed earlier for \textsc{ReducedOr} (Section~\ref{sec:operational}).
The optimised distribution used in $I_\text{broja}$ has directly coupled predictors:
\begin{align}
    \begin{split}
    P_\text{broja}(X_1=0,X_2=0) &= P_\text{broja}(X_1=1,X_2=1) = 0.5 \\
    P_\text{broja}(X_1=0,X_2=1) &= P_\text{broja}(X_1=1,X_2=0) = 0
    \end{split}
\end{align}
while the actual system has independent uniform marginal predictors ($P(i,j) = 0.25$).
In the $I_\text{broja}$ calculation of co-information the local events $(0,0,0)$ and $(1,1,2)$ both contribute redundant information, because $X_1$ and $X_2$ are coupled.
However, the local co-information terms for the true distribution show that the contributions of $x_1=0$ and $x_2=0$ are independent when $s=0$ (see Table~\ref{tab:pointwisesum}).
Therefore, with the true distribution these contributions are actually unique information. 
These differences arise because of the erroneous assumption within $I_\text{broja}$ that the unique and redundant information should be invariant to the predictor-predictor marginal distribution (Section~\ref{sec:operational}).
Since they are not, the $I_\text{broja}$ optimisation maximises redundancy by coupling the predictors.

The resulting $I_\text{ccs}$ PID seems quite intuitive.
Both $X_1$ and $X_2$ each tell whether the output sum is in $(0,1)$ or $(1,2)$, and they do this independently, since they are distributed independently (corresponding to $0.5$ bits of unique information each).
However, the final full discrimination of the output can only be obtained when both inputs are observed together, providing $0.5$ bits of synergy.
In contrast, $I_\text{broja}$ measures $0.5$ bits of redundancy.
It is hard to see how summation of two independent variables should be redundant as it is not apparent how two independent summands can convey overlapping information about their sum. 
For \textsc{and}, there is redundancy between two independent inputs.
$I_\text{ccs}$ shows that this arises from the fact that if $x_1=0$ then $y=0$ and similarly if $x_2=0$ then $y=0$. 
So when both $x_1$ and $x_2$ are zero they are both providing the same information content --- that $y=0$, so there is redundancy.
In contrast, in \textsc{sum}, $x_1=0$ tells that $y=0$ or $y=1$, but which of the two particular outputs is determined independently by the values of $x_2$.
So the information each input conveys is independent (unique) and not redundant.

\subsection{Griffith and Koch (2014) examples}
\label{sec:other2dexamples}

\textcite{griffith_quantifying_2014} present two other interesting examples: \textsc{RdnXor} (their Figure 6.9) and \textsc{RdnUnqXor} (their Figure 6.12).

\textsc{RdnXor} consists of two two-bit (4 value) inputs $X_1$ and $X_2$ and a two-bit (4 value) output $S$.
The first component of $X_1$ and $X_2$ redundantly specifies the first component of $S$.
The second component of $S$ is the \textsc{xor} of the second components of $X_1$ and $X_2$. 
This system therefore contains $1$ bit of redundant information and $1$ bit of synergistic information; further every value $s \in S$ has both a redundant and synergistic contribution.
$I_\text{ccs}$ correctly quantifies the redundancy and synergy with the PID $(1,0,0,1)$ (as do both $I_\text{min}$ and $I_\text{broja}$).

\textsc{RdnUnqXor} consists of two three-bit (8 value) inputs $X_1$ and $X_2$ and a four-bit (16 value) output $S$ (Figure~\ref{fig:rdnunqxor}).
The first component of $S$ is specified redundantly by the first components of $X_1$ and $X_2$. 
The second component of $S$ is specified uniquely by the second component of $X_1$ and the third component of $S$ is specified uniquely by the second component of $X_2$.
The fourth component of $S$ is the \textsc{xor} of the third components of $X_1$ and $X_2$.
Again $I_\text{ccs}$ correctly quantifies the properties of the system with the PID $(1,1,1,1)$, identifying the separate redundant, unique and synergistic contributions (as does $I_\text{broja}$ but not $I_\text{min}$).

Note that the PID with $I_\text{ccs}$ also gives the expected results for examples \textsc{Rnd} and \textsc{Unq} from \textcite{griffith_quantifying_2014}\footnote{See example scripts in the accompanying code.}.

\subsection{Dependence on predictor-predictor correlation}
\label{sec:predpred}

To directly illustrate the fundamental conceptual difference between $I_\text{ccs}$ and $I_\text{broja}$ we construct a family of distributions with the same target-predictor marginals and investigate the resulting decomposition as we change the predictor-predictor correlation \parencite{kay_finding_2017}.

We restrict our attention to binary variables with uniformly distributed univariate marginal distributions.
We consider pairwise marginals with a symmetric dependence of the form 
\begin{align}
\begin{split}
    p_c(0,0) &= p_c(1,1) = \nicefrac{(1+c)}{4} \\
    p_c(0,1) &= p_c(1,0) = \nicefrac{(1-c)}{4}
\end{split}
\end{align}
where the parameter $c$ specified the correlation between the two variables.
We fix $c=0.1$ for the two target-predictor marginals:
\begin{align}
\begin{split}
    P(X_1,S) &= P_{0.1}(X_1,S) \\
    P(X_2,S) &= P_{0.1}(X_2,S)
\end{split}
\end{align}
Then with $P(X_1,X_2) = P_c(X_1,X_2)$ we can construct a trivariate joint distribution $P_c(S,X_1,X_2)$ which is consistent with these three pairwise marginals as follows \parencite{kay_finding_2017}. 
This is a valid distribution for $-0.8 \leq c \leq 0.1$.
\begin{align}
\begin{split}
    p_c(0,0,0) &= \nicefrac{c}{4} + \nicefrac{1}{4} \\
    p_c(0,0,1) &= \nicefrac{1}{40} - \nicefrac{c}{4} \\
    p_c(0,1,0) &= \nicefrac{1}{40} - \nicefrac{c}{4} \\
    p_c(0,1,1) &= \nicefrac{c}{4} + \nicefrac{1}{5} \\
    p_c(1,0,0) &= 0 \\
    p_c(1,0,1) &= \nicefrac{9}{40} \\
    p_c(1,1,0) &= \nicefrac{9}{40} \\
    p_c(1,1,1) &= \nicefrac{1}{20}
    \label{eq:predpreddist}
\end{split}
\end{align}

\begin{figure}[htbp]
    \centering
    \includegraphics[width=0.9\textwidth]{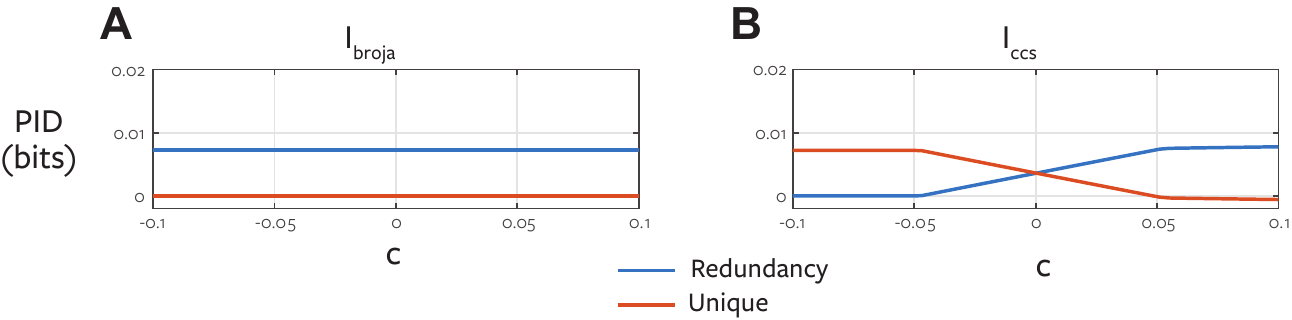}
    \caption{\emph{PIDs for binary systems with fixed target-predictor marginals as a function of predictor-predictor correlation.}
    $I_\text{broja}$ (A) and $I_\text{ccs}$ (B) PIDs are shown for the system defined in Eq.~\ref{eq:predpreddist} as a function of the predictor-predictor correlation $c$.}
    \label{fig:predpred}
\end{figure}

Figure~\ref{fig:predpred} shows $I_\text{broja}$ and $I_\text{ccs}$ PIDs for this system. 
By design the values of unique and redundant information obtained with $I_\text{broja}$ do not change as a function of predictor-predictor correlation when the target-predictor marginals are fixed.
With $I_\text{ccs}$ the quantities change in an intuitive manner. 
When the predictors are positively correlated, they are redundant, when they are negatively correlated they convey unique information.
When they are independent, there is an equal mix of unique and mechanistic redundancy in this system.
This emphasises the different perspective also revealed in the \textsc{ReducedOr} example (Section~\ref{sec:operational}) and the \textsc{and} example (Section~\ref{sec:and}).
$I_\text{broja}$ reports the co-information for a distribution where the predictors are perfectly coupled. 
For all the values of $c$ reported in Figure~\ref{fig:predpred}A, the $I_\text{broja}$ optimised distribution has coupled predictor-predictor marginals:
\begin{align}
\begin{split}
    P(X_1=0,X_2=1) &= P(X_1=1,X_2=0) = 0 \\
    P(X_1=0,X_2=0) &= P(X_1=1,X_2=1) = 0.5
\end{split}
\end{align}
Therefore, $I_\text{broja}$ is again insensitive to the sort of unique information that can be operationalised in a game-theoretic setting by exploiting the trial-by-trial relationships between predictors (Section~\ref{sec:operational}).

\section{Three variable examples}
\label{sec:threevarex}

We now consider the PID of the information conveyed about $S$ by three variables $X_1,X_2,X_3$.
For three variables we do not compare to $I_\text{broja}$, since it is defined only for two input sources.

\subsection{A problem with the three variable lattice?}
\label{sec:latticeproblem}

\textcite{bertschinger_shared_2013} identify a problem with the PID summation over the three-variable lattice (Figure \ref{fig:lattice}B).
They provide an example we term \textsc{XorCopy} (described in Sec.~\ref{sec:xorcopy}) which demonstrates that any redundancy measure satisfying their redundancy axioms (particularly the Harder et al.~identity axiom) cannot have only non-negative $I_\partial$ terms on the lattice.
We provide here an alternative example of the same problem, and one that does not depend on the particular redundancy measure used.
We argue it applies for any redundancy measure that attempts to measure overlapping information content.

\begin{figure}[htbp]
    \centering
    \includegraphics[width=0.6\textwidth]{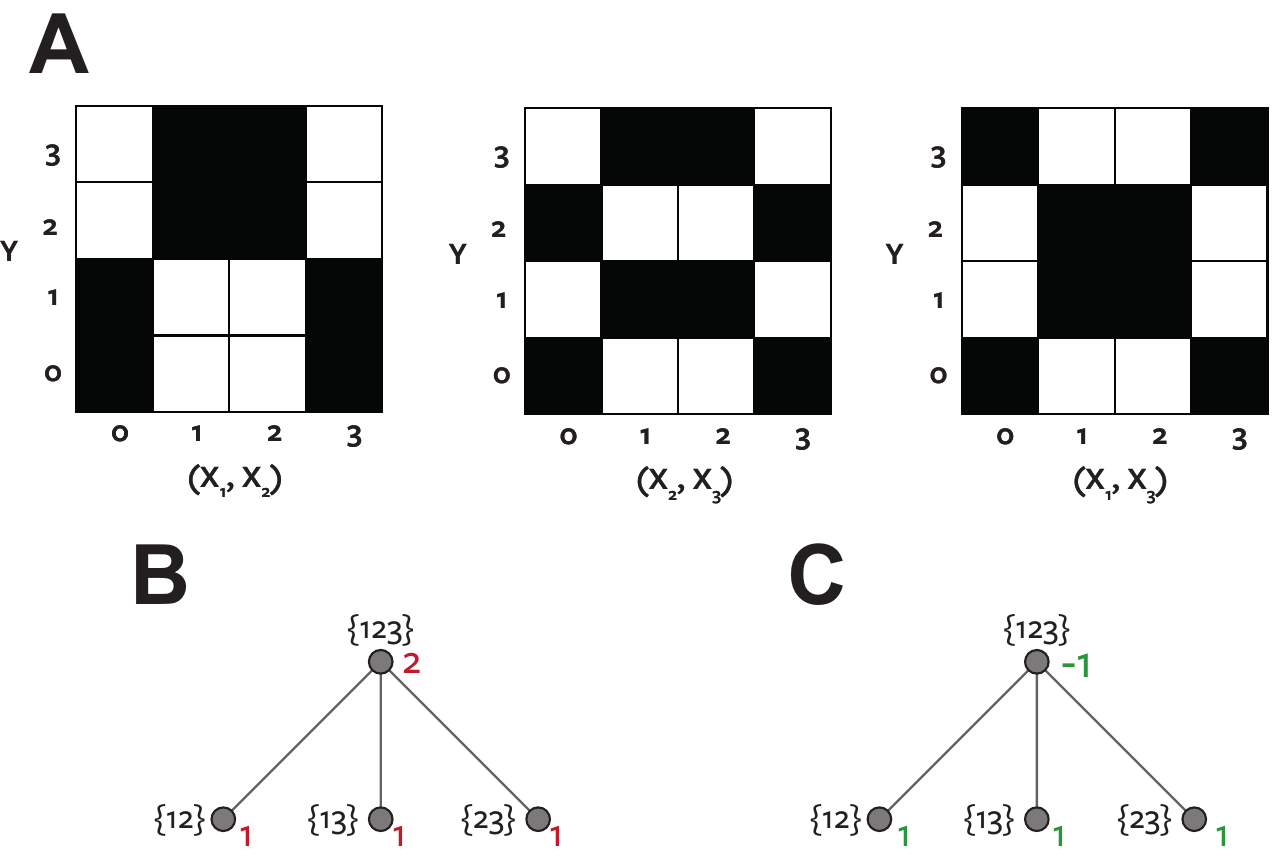}
    \caption{\emph{The \textsc{DblXor} example.}
    \textbf{A}: Pairwise variable joint distributions.
    Black tiles represent equiprobable outcomes.
    White tiles are zero-probability outcomes.
    \textbf{B}: Non-zero nodes of the three variable redundancy lattice. Mutual information values for each node are shown in red.
    \textbf{C}: PID. $I_\partial$ values for each node are shown in green.}
    \label{fig:dblxor}
\end{figure}

We consider $X_1,X_2,X_3$ independent binary input variables.
$Y$ is a two-bit (4 value) output with the first component given by $X_1 \oplus X_2$ and the second by $X_2 \oplus X_3$.
We refer to this example as \textsc{DblXor}.
In this case the top four nodes have non-zero (redundant) information: 
\begin{align*}
    I_\cap(\{123\})=I(\{123\}) &= 2 \mbox{ bits} \\
    I_\cap(\{12\})=I_\cap(\{13\})=I_\cap(\{23\}) &= 1 \mbox{ bit}
\end{align*}
We argue that all lower nodes on the lattice should have zero redundant (and partial) information.
First, by design and from the properties of \textsc{xor} no single variable conveys any information or can have any redundancy with any other source.
Second, considering synergistic pairs, Figure \ref{fig:dblxor}A graphically illustrates the source-output joint distributions for the two-variable sources.
Each value of the pairwise response (x-axes in Figure \ref{fig:dblxor}A) performs a different discrimination between the values of $Y$ for each pair. 
Therefore, there is no way there can be redundant information between any of these synergistic pairs. 
Redundant information means the same information content.
Since there are no discriminations (column patterns in the figure) that are common to more than one pair of sources, there can be no redundant information between them.
Therefore, the information conveyed by the three two-variable sources is also independent and all lower nodes on the lattice are zero.

In this example, $I_\cap(\{123\}) = 2$ but there are three child nodes of $\{123\}$ each with $I_\partial = 1$ (Figure \ref{fig:dblxor}B).
This leads to $I_\partial(\{123\})=-1$.
How can there be $3$ bits of unique information in the lattice when there are only $2$ bits of information in the system? 
In this case, we cannot appeal to the non-monotonicity of $I_\text{ccs}$ since these values are monotonic on the lattice.
There are also no negative pointwise terms in the calculation of $I(\{123\})$ so there is no synergistic misinformation that could explain a negative value.

In a previous version of this manuscript we argued that this problem arises because the three nodes in the penultimate level of the lattice are not disjoint, therefore not independent, and therefore mutual information is not additive over those nodes. 
We proposed a normalisation procedure to address such situations.
However, we now propose instead to accept the negative values. 
As noted earlier (Section~\ref{sec:properties}), negative values may subjectively be seen as a flaw \parencite{james_multivariate_2016}, but given that mutual information itself is a summation of positive and negative terms, there is no a priori reason why a full decomposition must, or indeed can, be completely non-negative.
In fact, in entropy terms, negative values are an essential consequence of the existence of mechanistic redundancy \parencite{ince_partial_2017}.
While in an information decomposition they can also arise from unique or synergistic misinformation, we propose that mechanistic redundancy is another explanation.
In this particular example of \textsc{DblXor}, the negative $\{123\}$ term reflects a mechanistic redundancy between the three pairwise synergistic partial information terms that cannot be accounted for elsewhere on the lattice. 

\subsection{Other three variable example systems}

\subsubsection{Giant bit and parity}

The most direct example of three-way information redundancy is the `giant bit' distribution \parencite{abdallah_measure_2012}.
This is the natural extension of example \textsc{rdn} (Section~\ref{sec:calculationexamples}) with a single bit in common to all four variables, defined as:
\begin{equation}
    P(0,0,0,0) = P(1,1,1,1) = 0.5
\end{equation}
Applying $I_\text{ccs}$ results in a PID with $I_\partial(S;\{1\}\{2\}\{3\})=1$ bit, and all other terms zero. 

A similarly classic example of synergy is the even parity distribution, a distribution in which an equal probability is assigned to all configurations with an even number of ones. 
The \textsc{xor} distribution is the even parity distribution in the three variable (two predictor) case.
Applying $I_\text{ccs}$ results in a PID with $I_\partial(S;\{123\})=1$ bit, and all other terms zero. 

Thus, the PID based on $I_\text{ccs}$ correctly reflects the structure of these simple examples.

\subsubsection{\textsc{XorCopy}}
\label{sec:xorcopy}

This example was developed to illustrate the problem with the three variable lattice described above \parencite{bertschinger_shared_2013,rauh_reconsidering_2014}.
The system comprises three binary input variables $X_1,X_2,X_3$, with $X_1,X_2$ uniform independent and $X_3=X_1 \oplus X_2$. 
The output $Y$ is a three bit (8 value) system formed by copying the inputs $Y=(X_1,X_2,X_3)$.
The PID with $I_\text{min}$ gives:
\begin{align*}
    I_\partial(\{1\}\{2\}\{3\}) = I_\partial(\{12\}\{13\}\{23\}) &= 1 \mbox{ bit}
\end{align*}
But since $X_1$ and $X_2$ are copied independently to the output it is hard to see how they can share information. 
Using common change in surprisal we obtain: 
\begin{align*}
    I_\text{ccs}(\{1\}\{23\})=I_\text{ccs}(\{2\}\{13\})=I_\text{ccs}(\{3\}\{12\}) &=1 \mbox{ bit} \\            
    I_\text{ccs}(\{12\}\{13\}\{23\}) &= 2 \mbox{ bits}
\end{align*}
The $I_\text{ccs}(\{i\}\{jk\})$ values correctly match the intuitive redundancy given the structure of the system, but result in a negative value similar to \textsc{DblXor} considered above.
There are $3$ bits of unique $I_\partial$ among the nodes of the third level, but only $2$ bits of information in the system.
This results in the PID:
\begin{align*}
    I_\partial(\{1\}\{23\})=I_\partial(\{2\}\{13\})=I_\partial(\{3\}\{12\}) &= 1 \mbox{ bit} \\            
    I_\partial(\{12\}\{13\}\{23\}) &= -1 \mbox{ bit}
\end{align*}
As for \textsc{DblXor} we believe this provides a meaningful decomposition of the total mutual information, with the negative value here representing the presence of mechanistic redundancy between the nodes at the third level of the lattice.
This mechanistic redundancy between synergistic pairs seems to be a signature property of an \textsc{xor} mechanism.

\subsubsection{Other examples}

\textcite{griffith_quantifying_2014} provide a number of other interesting three variable examples based on \textsc{xor} operations, such as \textsc{XorDuplicate} (their Figure 6.6), \textsc{XorLoses} (their Figure 6.7), \textsc{XorMultiCoal} (their Figure 6.14). 
For all of these examples $I_\text{ccs}$ provides a PID which matches what they suggest from the intuitive properties of the system (see \texttt{examples\_3d.m} in accompanying code).
$I_\text{ccs}$ also gives the correct PID for \textsc{ParityRdnRdn} (which appeared in an earlier version of their manuscript).

We propose an additional example, \textsc{XorUnq}, which consists of three independent input bits.
The output consists of 2 bits (4 values), the first of which is given by $X_1 \oplus X_2$, and the second of which is a copy of $X_3$. In this case we obtain the correct PID:
\begin{align*}
    I_\partial(\{3\}) = I_\partial(\{12\}) = 1 \mbox{ bit}
\end{align*}

\begin{figure}[htbp]
    \centering
    \includegraphics[width=0.6\textwidth]{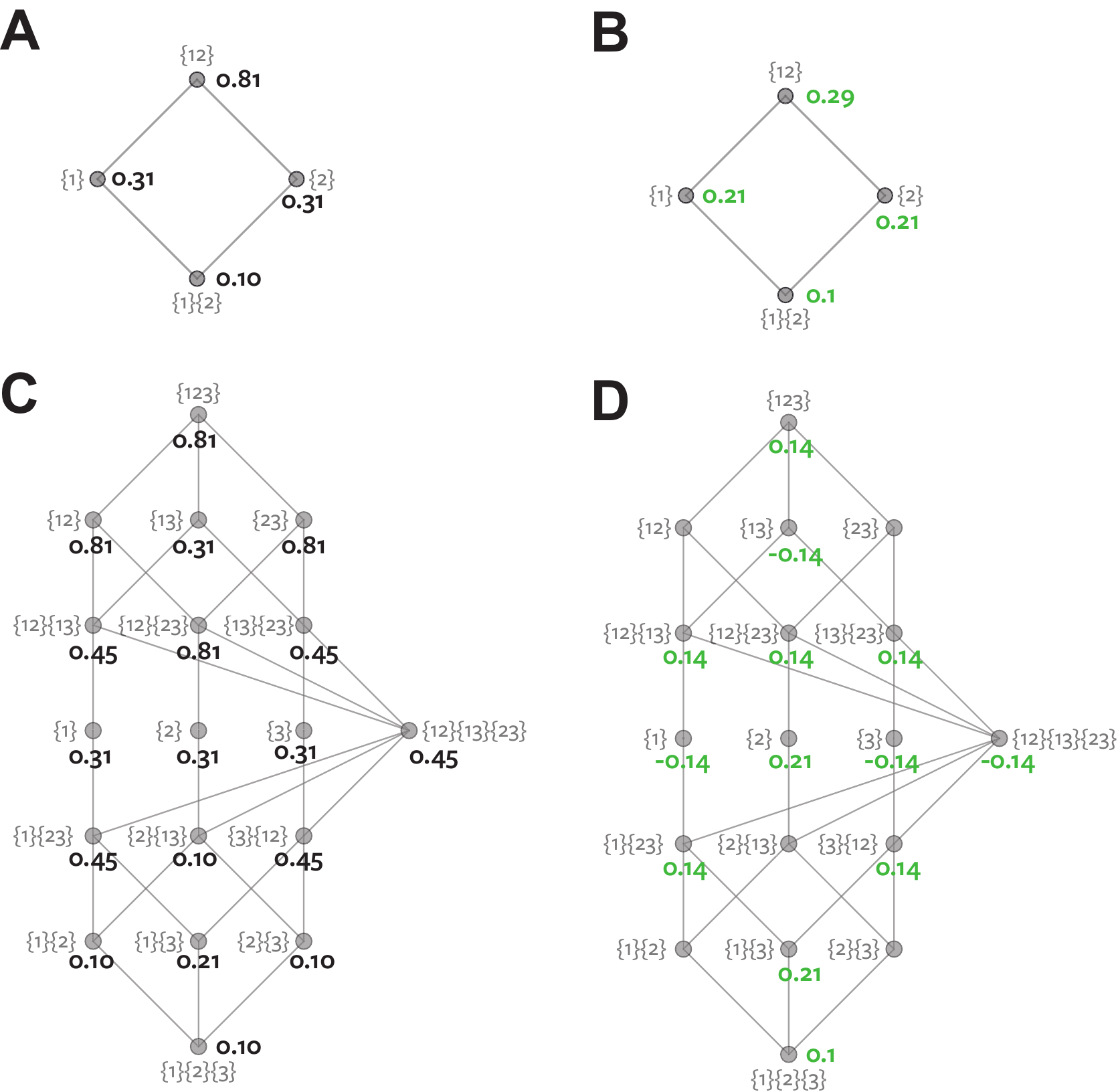}
    \caption{\emph{The \textsc{AndDuplicate} example.}
    \textbf{A}: $I_\text{ccs}$ values for \textsc{and}.
    \textbf{B}: Partial information values from the $I_\text{ccs}$ PID for \textsc{and}.
    \textbf{C}: $I_\text{ccs}$ values for \textsc{AndDuplicate}.
    \textbf{D}: Partial information values from the $I_\text{ccs}$ PID for \textsc{AndDuplicate}.
    }
    \label{fig:andduplicate}
\end{figure}

Another interesting example from \textcite{griffith_quantifying_2014}  is \textsc{AndDuplicate} (their Figure 6.13). 
In this example $Y$ is a binary variable resulting from the binary \textsc{and} of $X_1$ and $X_2$.
$X_3$ is a duplicate of $X_1$.
The PID we obtain for this system is shown in Figure~\ref{fig:andduplicate}.

We can see that as suggested by \textcite{griffith_quantifying_2014}, 
\begin{align}
    \begin{split}
        I_\partial^\text{\textsc{AndDup}}(S;\{2\}) &= I_\partial^\text{\textsc{and}}(S;\{2\})\\
        I_\partial^\text{\textsc{AndDup}}(S;\{1\}\{3\}) &= I_\partial^\text{\textsc{and}}(S;\{1\})\\
        I_\partial^\text{\textsc{AndDup}}(S;\{1\}\{2\}\{3\}) &= I_\partial^\text{\textsc{and}}(S;\{1\}\{2\})
    \end{split}
\end{align}

The synergy relationship they propose, $I_\partial^\text{\textsc{AndDup}}(S;\{12\}\{23\}) = I_\partial^\text{\textsc{and}}(S;\{12\})$ is not met, although the fundamental general consistency requirement relating 2 and 3 variable lattices is \parencite{chicharro_synergy_2017,ince_partial_2017}:
\begin{align}
\begin{split}
I_\partial^\text{\textsc{and}}(S;\{12\}) &= I_\partial^\text{\textsc{AndDup}}(S;\{12\})  \\
&+ I_\partial^\text{\textsc{AndDup}}(S;\{12\}\{13\}) + I_\partial^\text{\textsc{AndDup}}(S;\{12\}\{23\}) \\
&+ I_\partial^\text{\textsc{AndDup}}(S;\{12\}\{13\}\{23\}) \\
&+I_\partial^\text{\textsc{AndDup}}(S;\{3\}\{12\}) 
\end{split}
\end{align}

Note that the preponderance of positive and negative terms with amplitude $0.14$ bits is at first glance counter-intuitive, particularly the fact that $I_\partial^\text{\textsc{AndDup}}(S;\{1\}) = I_\partial^\text{\textsc{AndDup}}(S;\{3\})=-0.146$ when $X_3$ is a copy of $X_1$.
However, the $0.14$ bits comes from a local misinformation term in the univariate predictor-target mutual information calculation for \textsc{and}, which is not present in the joint mutual information calculation.
This reflects the fact that, in entropy terms, $I(S;X_1)$ is not a proper subset of $I(S;X_1, X_2)$ \parencite{ince_partial_2017}.
A partial entropy decomposition of \textsc{and} shows that $H_\partial(\{1\}\{23\}) = H_\partial(\{2\}\{13\}) = 0.14$.
These are entropy terms that have an ambiguous interpretation and appear both in unique and synergistic partial information terms. 
It is likely that a higher-order entropy decomposition could shed more light on the structure of the \textsc{AndDuplicate} PID.

\section{Continuous Gaussian Variables}

$I_\text{ccs}$ can be applied directly to continuous variables.
$\Delta_s h^{\text{com}}$ can be used locally in the same way, with numerical integration applied to obtain the expectation\footnote{Functions implementing this for Gaussian variables via Monte Carlo integration are included in the accompanying code.}.
Following \textcite{barrett_exploration_2015} we consider the information conveyed by two Gaussian variables $X_1,X_2$ about a third Gaussian variable, $S$.
We focus here on univariate Gaussians, but the accompanying implementation also supports multivariate normal distributions.
\textcite{barrett_exploration_2015} show that for such Gaussian systems, all previous redundancy measures agree, and are equal to the minimum mutual information carried by the individual variables:
\begin{align}
    I_\cap(\{1\}\{2\}) = \min_{i=1,2} I(S; X_i) = I_\text{mmi}(\{1\}\{2\})
\end{align}

\begin{figure}[htbp]
    \centering
    \includegraphics[width=0.8\textwidth]{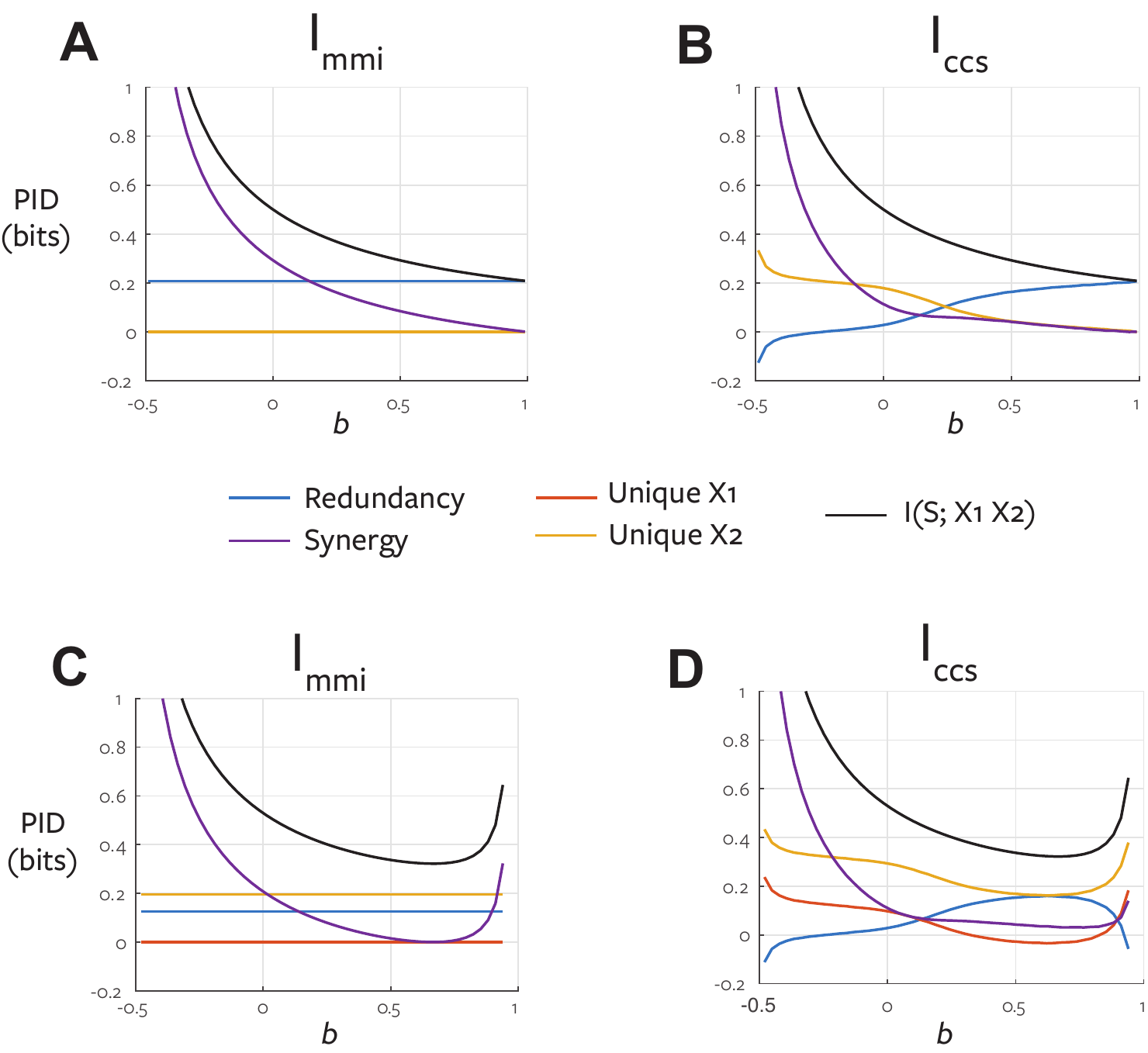}
    \caption{\emph{PIDs for Gaussian systems.}
    \textbf{A}: PID with $I_\text{mmi}$ for $a=c=0.5$ as a function of predictor-predictor correlation $b$.
    \textbf{B}: PID with $I_\text{ccs}$ for $a=c=0.5$.
    \textbf{C}: PID with $I_\text{mmi}$ for $a=0.4,c=0.6$.
    \textbf{D}: PID with $I_\text{ccs}$ for $a=0.4,c=0.6$.
     }
    \label{fig:gauss}
\end{figure}

Without loss of generality, we consider all variables to have unit variance, and the system is then completely specified by three parameters:
\begin{align*}
    a &= \text{Corr}(X_1, S) \\
    c &= \text{Corr}(X_2, S) \\
    b &= \text{Corr}(X_1, X_2)
\end{align*}
Figure \ref{fig:gauss} shows the results for two families of Gaussian systems as a function of the correlation, $b$, between $X_1$ and $X_2$ \parencite[][Figure 3]{barrett_exploration_2015}.

This illustrates again a key conceptual difference between $I_\text{ccs}$ and existing measures. 
$I_\text{ccs}$ is not invariant to the predictor-predictor marginal distributions (Section~\ref{sec:predpred}).
When the two predictors have equal positive correlation with the target (Figure~\ref{fig:gauss}A,B), $I_\text{mmi}$ reports zero unique information, and a constant level of redundancy regardless of the predictor-predictor correlation $b$.
$I_\text{ccs}$ transitions from having the univariate predictor information purely unique when the predictors are negatively correlated, to purely redundant when the predictors are strongly positively correlated.
When the two predictors have unequal positive correlations with the target (Figure~\ref{fig:gauss}C,D), the same behaviour is seen.
When the predictors are negatively correlated the univariate information is unique, as they become correlated both unique informations decrease as the redundancy between the predictors increases. 

Having an implementation for continuous Gaussian variables is of practical importance, because for multivariate discrete systems sampling high dimensional spaces with experimental data becomes increasingly challenging.
We recently developed a lower-bound approximate estimator of mutual information for continuous signals based on a Gaussian copula \parencite{ince_statistical_2017}.
The Gaussian $I_\text{ccs}$ measure therefore allows this approach to be used to obtain PIDs from experimental data. 

\section{Discussion}

We have presented $I_\text{ccs}$, a novel measure of redundant information based on the expected pointwise change in surprisal that is common to all input sources.
Developing a meaningful quantification of redundant and synergistic information has proved challenging, with disagreement about even the basic axioms and properties such a measure should satisfy.
Therefore, here we take a bottom-up approach, starting by defining what we think redundancy should measure at the pointwise level (common change in surprisal), and then exploring the consequences of this through a range of examples.

This new redundancy measure has several advantages over existing proposals.
It is conceptually simple: it measures precisely the pointwise contributions to the mutual information which are shared unambiguously among the considered sources.
This seems a close match to an intuitive definition of redundant information.
$I_\text{ccs}$ exploits the additivity of surprisal to directly measure the pointwise overlap as a set intersection, while removing the ambiguities that arise due to the conflation of pointwise information and misinformation effects by considering only terms with common sign (since a common sign is a prerequisite for there to be a common change in surprisal).
$I_\text{ccs}$ is defined for any number of input sources (implemented for 2 and 3 predictor systems), as well as any continuous system (implemented for multivariate Gaussian predictors and targets).
Matlab code implementing the measure accompanies this article\footnote{%
Available at: \url{https://github.com/robince/partial-info-decomp} \newline
Requires installation of Python and the dit toolbox \parencite{james_dit/dit_2017}.}.
The repository includes all the examples described herein, and it is straightforward for users to apply the method to any other systems or examples they would like.

To motivate the choice of joint distribution we use to calculate $I_\text{ccs}$ we review and extend the decision theoretic operational argument of \textcite{bertschinger_quantifying_2014}.
We show how a game theoretic operationalisation provides a different perspective, and give a specific example where an exploitable game-theoretic advantage exists for each agent, but  $I_\text{broja}$ suggests there should be no unique information.
We therefore conclude the decision theoretic formulation is too restrictive and that the balance of unique and redundant information is not invariant to changes in the predictor-predictor marginal distribution.  
This means that the optimisation in $I_\text{broja}$ is not only minimising synergy, but could actually be increasing redundancy.
Detailed consideration of several examples shows that the $I_\text{broja}$ optimisation often results in distributions with coupled predictor variables, which maximises the source redundancy between them.
For example, in the \textsc{sum} system, the coupled predictors make the $(0,0,0)$ and $(1,1,2)$ events redundant, when in the true system the predictors are independent, so those events contribute unique information. 
However, we note that if required $I_\text{ccs}$ can also be calculated following the decision theoretic perspective simply by using $\hat{P}_\text{ind}$. 

$I_\text{ccs}$ satisfies most of the core axioms for a redundancy measure, namely symmetry, self-redundancy and a modified identity property which reflects the fact that mutual information can itself include synergistic entropy effects \parencite{ince_partial_2017}.
Crucially, it also satisfies subset equality which has not previously been considered separately from monotonicity, but is the key axiom which allows the use of the reduced redundancy lattice.
However, we have shown that $I_\text{ccs}$ is not monotonic on the redundancy lattice because nodes can convey unique misinformation.
This means the resulting PID is not non-negative.
In fact, negative terms can occur even without non-monotonicity because for some systems (e.g. 3 predictor systems with \textsc{xor} structures) mechanistic redundancy can result in negative terms \parencite{ince_partial_2017}.
We argue that while ``negative \dots~atoms can subjectively be seen as flaw''  \parencite{james_multivariate_2016} in fact, they are a necessary consequence of a redundancy measure that genuinely quantifies overlapping information content.
We have shown that despite the negative values, $I_\text{ccs}$ provides intuitive and consistent PIDs across a range of example systems drawn from the literature.

Mutual information itself is an expectation over positive and negative terms.
While Jensen's inequality ensures that the overall expectation is non-negative, we argue there is no way to apply Jensen's inequality to decomposed partial information components of mutual information, whichever redundancy measure is used, and thus no reason to assume they must be non-negative.
An alternative way to think about the negative values is to consider the positive and negative contributions to mutual information separately. 
The definition of $I_\text{ccs}$ could easily be expanded to quantify redundant pointwise information separately from redundant pointwise misinformation (rows 1 and 3 of Table~\ref{tab:intsign}).
One could then imagine two separate lattice decompositions, one for the pointwise information (positive terms) and one for the pointwise misinformation (negative terms).
We conjecture that both of these lattices would be monotonic, and that the non-monotonicity of the $I_\text{ccs}$ PID arises as a net effect from taking the difference between these.
This suggests it may be possible to obtain zero unique information from a cancellation of redundant information with redundant misinformation, analogous to how zero co-information can result in the presence of balanced redundant and synergistic effects, and so exploring this approach is an interesting area for future work.
It is also important to develop more formal analytical results proving further properties of the measure, and separate local information versus local misinformation lattices might help with this.

\textcite{rauh_secret_2017} recently explored an interesting link between the PID framework and the problem of cryptographic secret sharing. 
Intuitively, there should be a direct relationship between the two notions: an authorized set should have only fully synergistic information about the secret when all elements of the set are considered, and a shared secret scheme corresponds to redundant information about the secret between the authorized sets. 
Therefore, any shared secret scheme should yield a PID with a single non-negative partial information term equal to the entropy of the secret at the node representing the redundancy between the synergistic combinations of each authorised set within the inclusion-minimal access structure.
\textcite{rauh_secret_2017} shows that if this intuitive relationship holds, then the PID cannot be non-negative. 
This finding further supports our suggestion that it may not be possible to obtain a non-negative PID from a redundancy measure that meaningfully quantifies overlapping information content; if such a measure satisfies the intuitive `secret sharing property' \parencite{rauh_secret_2017} it does not provide a non-negative PID.
We note that $I_\text{ccs}$ satisfies  the secret sharing property for \parencite[Example 1]{rauh_secret_2017}; whether it can be proved to do so in general is an interesting question for future research.
These considerations suggest $I_\text{ccs}$ might be useful in cryptographic applications. 

Another important consideration for future research is how to address the practical problems of limited sampling bias \parencite{panzeri_correcting_2007} when estimating PID quantities from experimental data. 
Similarly, how best to perform statistical inference with non-parametric permutation methods is an open question.
We suggest it is likely that different permutation schemes might be needed for the different PID terms, since trivariate conditional mutual information requires a different permutation scheme than bivariate joint mutual information \parencite{ince_novel_2012}.

How best to practically apply the PID to systems with more than three variables is also an important area for future research.
The four variable redundancy lattice has $166$ nodes, which already presents a significant challenge for interpretation if there are more than a handful of non-zero partial information values.  
We suggest that it might be useful to collapse together the sets of terms that have the same order structure.
For example, for the three variable lattice the terms within the layers could be represented as shown in Table \ref{tab:reducedlattice}.
While this obviously does not give the complete picture provided by the full PID, it gives considerably more detail than existing measures based on maximum entropy subject to different order marginal constraints, such as connected information \parencite{schneidman_network_2003}.
We hope it might provide a more tractable practical tool that can still give important insight into the structure of interactions for systems with four or more variables.

\begin{table}[htbp]
\begin{center}
    \begin{tabular}{| c  | c | c | c | c |}
        \hline
        \textbf{Level} & \textbf{Order-structure terms} & \textbf{Giant bit} & \textbf{Parity} & \textbf{\textsc{DblXor}} \\
        \hline \hline
        7              & $(3)$                         & 0 & 1 & -1                       \\
        6              & $(2)$                         & 0 & 0 & 3                        \\
        5              & $(2,2)$                       & 0 & 0 & 0                        \\
        4              & $(1)$, $(2,2,2)$              & 0,0 & 0,0 & 0, 0                     \\
        3              & $(1,2)$                       & 0 & 0 & 0                        \\
        2              & $(1,1)$                       & 0 & 0  & 0                        \\
        1              & $(1,1,1)$                     & 1 & 0  & 0                        \\
        \hline
        \end{tabular}
        \caption{\emph{Order-structure terms for the three variable lattice}. Resulting values for the example systems of a giant bit, even parity and \textsc{DblXor} (Section~\ref{sec:threevarex}) are shown.}
  \label{tab:reducedlattice}
\end{center}
\end{table}

We have recently suggested that the concepts of redundancy and synergy apply just as naturally to entropy as to mutual information \parencite{ince_partial_2017}.
Therefore, the redundancy lattice and PID framework can be applied to entropy to obtain a partial entropy decomposition.
A particular advantage of the entropy approach is that it provides a way to separately quantify source and mechanistic redundancy \parencite{harder_bivariate_2013,ince_partial_2017}.
Just as mutual information is derived from differences in entropies, we suggest that partial information terms should be related to partial entropy terms.
For any partial information decomposition, there should be a compatible partial entropy decomposition. 
We note that $I_\text{ccs}$ is highly consistent with a PID based on a partial entropy decomposition obtained with a pointwise entropy redundancy measure which measures common surprisal \parencite{ince_partial_2017}. 
More formal study of the relationships between the two approaches is an important area for future work.
In contrast, it is hard to imagine an entropy decomposition compatible with $I_\text{broja}$.
In fact, we have shown that $I_\text{broja}$ is fundamentally incompatible with the notion of synergistic entropy.
Since it satisfies the Harder et al.~identity axiom, it induces a two variable entropy decomposition which always has zero synergistic entropy.

As well as providing the foundation for the PID, a conceptually well-founded and practically accessible measure of redundancy is a useful statistical tool in its own right.
Even in the relatively simple case of two experimental dependent variables, a rigorous measure of redundancy can provide insights about the system that would not be possible to obtain with classical statistics. 
The presence of high redundancy could indicate a common mechanism is responsible for both sets of observations, whereas independence would suggest different mechanisms.
To our knowledge the only established approaches that attempt to address such questions in practice are Representational Similarity Analysis \parencite{kriegeskorte_representational_2008} and cross-decoding methods such as the temporal generalisation method \parencite{king_characterizing_2014}.
However, both these approaches can be complicated to implement, have restricted domains of applicability and cannot address synergistic interactions. 
We hope the methods presented here will provide a useful and accessible alternative allowing statistical analyses that provide novel interpretations across a range of fields.

\section*{Acknowledgements}

I thank Jim Kay, Michael Wibral, Ryan James, Johannes Rauh and Joseph Lizier for useful discussions.
I thank Ryan James for producing and maintaining the excellent dit package.
I thank Daniel Chicharro for introducing me to the topic and providing many patient explanations and examples. 
I thank Eugenio Piasini, Christoph Daube and Philippe Schyns for useful comments on the manuscript.

\printbibliography

\end{document}